\documentclass[journal,draftcls,onecolumn,12pt,twoside]{IEEEtran}

\usepackage[utf8]{inputenc} 
\usepackage[T1]{fontenc}
\usepackage{url}
\usepackage{ifthen}
\usepackage{cite,color}
\usepackage{amsmath}
\usepackage{amssymb}
\usepackage{graphicx}
\usepackage{comment}
\newtheorem{thm}{Theorem}
\newtheorem{rk}{Remark}
\newtheorem{cor}{Corollary}
\newtheorem{de}{Definition}
\newtheorem{pro}{Proposition}
\usepackage{subcaption} % or use \usepackage{subfigure} for older LaTeX versions

\usepackage{flushend} % to balance last page
\bstctlcite{IEEEexample:BSTcontrol} % to balance last page
\begin{document}
\title{Detection of Signals in Colored Noise: Roy's Largest Root Test for Non-central $F$-matrices } 

\author{
	\vspace*{-0mm}
	Prathapasinghe~Dharmawansa,~Saman~Atapattu,~\IEEEmembership{Senior Member, IEEE},\\ Jamie~Evans,~\IEEEmembership{Senior~Member,~IEEE},~and M\'erouane~Debbah,~\IEEEmembership{Fellow,~IEEE}
	% \thanks{}
  \thanks{This work was presented in part at the IEEE International Symposium on Information Theory (ISIT 2024), Athens, Greece, Jul. 2024 \cite{Dharmawansa2024ISIT}.}
 \thanks{P. Dharmawansa and J. Evans are with the Department of Electrical and Electronic Engineering, University of Melbourne, Parkville, VIC 3010, Australia (e-mail: \{prathapa.d, jse\}@unimelb.edu.au).}%
		\thanks{S. Atapattu is with the Department of Electrical and Electronic Engineering, School of Engineering, RMIT University, Melbourne, VIC, Australia (e-mail: saman.atapattu@rmit.edu.au).}
  \thanks{M. Debbah is with KU 6G Research Center, Khalifa University of Science and Technology, P O Box 127788, Abu Dhabi, UAE (email: merouane.debbah@ku.ac.ae) and also with CentraleSupelec, University Paris-Saclay, 91192 Gif-sur-Yvette, France.}
\vspace{-5mm}  
}

\maketitle

\begin{abstract}
   This paper investigates the signal detection problem in colored noise with an unknown covariance matrix. In particular, we focus on detecting a non-random signal by capitalizing on the leading eigenvalue (a.k.a. Roy's largest root) of the whitened sample covariance matrix as the test statistic. To this end, the whitened sample covariance matrix is  constructed via \(m\)-dimensional \(p \) plausible signal-bearing samples and \(m\)-dimensional \(n \) noise-only samples. Since the signal is non-random, the whitened sample covariance matrix turns out to have a {\it non-central} \(F\)-distribution with a rank-one non-centrality parameter. Therefore, the performance of the test entails the statistical characterization of the leading eigenvalue of the non-central \(F\)-matrix, which we address by deriving its cumulative distribution function (c.d.f.) in closed-form by leveraging the powerful orthogonal polynomial approach in random matrix theory. This new c.d.f. has  been instrumental in analyzing the receiver operating characteristic (ROC) of the detector. We also extend our analysis into the high dimensional regime in which \(m,n\), and \(p\) diverge such that \(m/n\) and \(m/p\) remain fixed. It turns out that, when \(m=n\) and fixed, the power of the test improves if the signal-to-noise ratio (SNR) is of at least \(O(p)\), whereas the corresponding SNR in the high dimensional regime is of at least \(O(p^2)\). Nevertheless, more intriguingly, for \(m<n\) with the SNR  of order \(O(p)\), the leading eigenvalue does not have power to detect {\it weak} signals in the high dimensional regime.      
\end{abstract}

% \newpage

\section{Introduction}
The fundamental problem of detecting a signal embedded in noise has been at the forefront of various research studies, see e.g., \cite{nadakuditi,dissanayake2022,chamain2020eigenvalue,debbah2011,edelman,mcwhorter2023,johnstone2020,YCliang,larson,hemimo,couillet} and references therein. In this regard, among various detectors available, a certain class of detectors rely exclusively on specific correlation or mean structures inherent in the observational data \cite{nadakuditi,dissanayake2022,chamain2020eigenvalue,johnstone2020,YCliang,larson}. For instance, finite rank perturbation of the identity covariance matrix (i.e., low-rank-plus-identity) and rank deficient mean matrix are two such prominent structures \cite{johnstone2001,larson,chamain2020eigenvalue,debbah2011,nadakuditi,johnstone2020}.
These key structures are the consequences of certain specific signal/channel characteristics which may or may not be known at the receiver.
The detectors which exploit those covariance or mean structures without the knowledge of underlying  signal/channel characteristics are referred to as  blind detectors \cite{YCliang,larson}. It turns out that those key structural characteristics are embedded in the eigenvalues of the corresponding matrices. Therefore, certain functions of the eigenvalues of 
the covariance matrix have been used as the test statistics in such detectors.

Notwithstanding the above facts, in practice, since the exact covariance matrix cannot be computed,  the covariance matrix is estimated with the available data samples. Therefore, corresponding test statics are also computed based on the sample eigenvalues, instead.  In this regard, among various eigenvalue based test statistics available, the leading eigenvalue of the sample covariance matrix, which is also known as Roy's largest root\footnote{The Roy's largest root test is a direct consequence of Roy's union-intersection principle \cite{mardia2024}.}, has been a popular choice among detection theorists \cite{Kritchman,johnstoneRoy,chamain2020eigenvalue,prathapRoyRoot,wang2017stat,YCliang,larson} due to its certain optimal properties. To be specific, the largest
root test turns out to be most powerful among the common tests when
the alternative is of rank-one \cite{johnstoneRoy,Kritchman}. The rank-one alternative manifests as a single {\it spike} with respect to either the covariance or non-centrality parameter matrices \cite{johnstone2001,baikPhase,johnstoneRoy,Dharmawansa2014}. Here our interest is also in a rank-one none-centrality parameter alternative.

The most classical additive noise abstraction is the white Gaussian noise. However, in most modern practical settings, the additive Gaussian noise turns out to have certain covariance structures \cite{chamain2020eigenvalue,richmond,hemimo,mcwhorter2023,Vinogradova,werner2007,werner2006,dogandzic2003generalized,liu2018bayesian,rong2022adaptive} , thereby it is referred to as colored noise. For instance, in applications pertaining to radar detection, the effective noise is commonly modeled as colored, in particular, to account for thermal noise, jamming, and clutter effects \cite{richmond,hemimo,mcwhorter2023,werner2007,werner2006,dogandzic2003generalized,liu2018bayesian,rong2022adaptive}. The highly unlikely scenario of known noise covariance matrix at the receiver, the detection theorist prefers to work with respect to a rotated and scaled coordinate system in which the effective noise is white. This process is known as whitening. However, as we have already mentioned in an early instance, in practice, the noise covariance is unknown and therefore, should be estimated. To facilitate the computation of the noise sample covariance matrix, the assumption of the availability of noise-only data (i.e., signal-free data or secondary data) has been introduced in the literature \cite{dogandzic2003generalized,nadakuditi,chamain2020eigenvalue,johnstoneRoy,richmond,werner2007,zachariah,melvin2004,rong2022adaptive}. To be specific, here we assume that the detector is made available with a separate set of noise-only samples in addition to plausible signal-bearing samples (i.e., primary data).
Consequently, the whitening operation can now be conveniently applied to the primary data to obtain the white noise equivalent primary data.

Here our focus is on detecting a non-random signal in colored noise with an unknown covariance matrix by capitalizing on the underlying specific covariance/mean structure of the observations. Particular examples include, detection problems associated with modern multiple-input-multiple-output (MIMO) radar and classical radar sensing applications (see e.g., \cite{tang2020polyphase,liu2018bayesian,cui2013,li2007mimo,haimovich2007mimo,fishler2006,dogandzic2003generalized,rong2022,besson2023} and references therein). To leverage the potential of underpinning correlation/mean structure of the observations, we focus on the whitened sample covariance matrix. To be specific, assuming the availability of primary (i.e., plausible signal bearing samples)  and secondary data (i.e., noise-only samples) sets, the whitened sample covariance matrix can conveniently be written in its symmetric form as $\widehat{\boldsymbol{F}}=\widehat{\boldsymbol{\Sigma}}^{-1/2}\boldsymbol{\widehat R}\widehat{\boldsymbol{\Sigma}}^{-1/2}$ where  $\widehat{\boldsymbol{\Sigma}}\in\mathbb{C}^{m\times m}$ is the estimated noise covariance matrix, $\widehat{\boldsymbol{R}}\in\mathbb{C}^{m\times m}$ denotes the sample covariance estimated with the primary data, $m$ is the system dimensionality and $(\cdot)^{1/2}$ denotes the positive definite square root. Since the signal of our interest is non-random, under the Gaussian noise assumption, $\widehat{\boldsymbol{R}}\in\mathbb{C}^{m\times m}$ turns out to have a non-central Wishart distribution with a rank-one non-centrality parameter, whereas $\widehat{\boldsymbol{\Sigma}}\in\mathbb{C}^{m\times m}$ has a central Wishart distribution. Consequently,  $\widehat{\boldsymbol{F}}$ follows  the so-called non-central $F$-distribution \cite{james,muirhead,johnstoneRoy} with a rank-one non-centrality. In certain situations, the number of available samples may be few in comparison to the system dimension $m$ (i.e., sample deficiency) \cite{Vinogradova,vallet2017performance,pham2015,mestre2008}. In such situation, the matrix $\widehat{\boldsymbol{F}}$ becomes rank deficient, there by having a singular non-central $F$-distribution \cite{james}. Since the non-centrality parameter of $F$-matrix is of rank-one, having motivated with certain optimal properties of the leading eigenvalue pertaining to rank-one alternatives, here we employ the leading eigenvalue of the $F$-matrix as the test statistic.  The utility of the leading eigenvalue based test is further highlighted, as delineated in \cite{johnstoneRoy,muirhead}, by the fact that the Roy's largest root test is most powerful among the common tests when the alternative is of rank-one \cite{Kritchman}. Therefore, we are interested in the statistical characteristics of the leading eigenvalue of non-central $F$-matrix with a rank-one non-centrality parameter. Some preliminary limited analyses in this respect have recently appeared in \cite{Dharmawansa2024ISIT}. 

The joint eigenvalue densities of non-central singular and non-singular $F$-matrices have been reported in \cite{james}. A stochastic representation of the leading eigenvalue of  a non-central $F$-matrix with a rank-one non-centrality parameter in the high signal-to-noise ratio (SNR) regime  is given in \cite{johnstoneRoy,prathapRoyRoot}. 
The high dimensional statistical characteristics of the eigenvalues of central $F$-matrices have been well documented in the literature (see e.g., \cite{nadakuditi,wang2017stat,Johnstone2008,jiang2022invariance,han2016tracy,Guan, jiang2021,WangYao} and references therein). Nevertheless, only a few results are available for the non-central $F$-matrices \cite{Dharmawansa2014,johnstone2020,hou2023spiked,Guan}. As delineated in \cite{Dharmawansa2014}, in the presence of the so-called non-central spikes (i.e., the non-zero eigenvalues of the non-centrality parameter matrix), the eigenvalues of non-central $F$-matrices undergo {\it phase transition\footnote{This phenomenon is commonly known as the Baik, Ben Arous, P\'ech\'e (BBP) phase transition because of their seminal contribution in \cite{ref:baikPhaseTrans}. The signal processing analogy of this phenomenon is known as the ``subspace swap" \cite{ref:johnsonMestre,ref:thomas,ref:tuft}.}}, the threshold of which is also derived therein. To be specific, when the population non-central spikes  are below the phase transition threshold (i.e., in the sub-critical regime), the corresponding sample eigenvalues of a non-central $F$-matrix converges to the right edge of the bulk spectrum and satisfy the most celebrated Tracy-Widom law \cite{Guan}, whereas when the population non-central spikes are above the phase transition threshold (i.e., in the super-critical regime), the corresponding sample eigenvalues follow a joint Gaussian density \cite{Dharmawansa2014,hou2023spiked}. Nevertheless, the above high dimensional results do not hold in the special case of an equality between the number of noise-only samples and the system dimensionality. Moreover, a tractable finite-dimensional characterization of the leading eigenvalue of a non-central $F$-matrix is also not reported in the current literature.

Having motivated with the above facts and noting the limited very recent results in \cite{Dharmawansa2024ISIT}, in this paper, capitalizing on powerful orthogonal polynomials technique advocated in random matrix theory \cite{mehta}, we derive a new exact cumulative distribution function (c.d.f.) for the leading eigenvalue of a non-central $F$-matrix with a rank-one non-centrality parameter matrix. 
The new c.d.f. expression consists of a determinant of a square matrix whose dimension depend on the relative difference between the number of noise only samples $n$ and the system dimensionality $m$ (i.e., $n-m$) but not their individual magnitudes. This key feature further facilitates the efficient evaluation of the c.d.f. corresponding to an important configuration $n=m$ (i.e., when the noise sample covariance matrix is nearly rank deficient). Since the parameter $n-m$ can also be considered as an implicit indicator of the quality of $\widehat{\boldsymbol{\Sigma}}$ as an estimator of the unknown population noise covariance matrix, the above configuration corresponds to the lowest quality noise covariance estimator. Therefore, this configuration in turn dictates a performance lower bound on the leading eigenvalue as a test statistic for other parameter being fixed. This new c.d.f. expression further facilitates the analysis of the receiver operating characteristics (ROC) of the largest root test. Moreover, 
 driven by the modern high dimensional statistical applications involving the non-central $F$-matrices \cite{johnstone2020,hou2023spiked,Guan}, we have extended our analysis to the high dimensional regime in which $m,n,p\to\infty$ such that $m/p\to c_1\in(0,1)$ and $m/n\to c_2\in(0,1]$, where $p$ is the number of plausible signal-bearing samples.

 The key analytical results developed in this paper shed some light on the impact of the  the system dimension (i.e., $m$), the number of plausible signal-bearing samples  (i.e., $p$), noise-only samples (i.e., $n$), and signal-to-noise ratio (SNR) (i.e., $\gamma$) on the ROC.
For instance, our analytical ROC results corresponding to the scenario, for which $m=n$ (i.e., the number of noise-only samples equals the system dimensionality) and fixed, reveal that the power is an increasing function of $p$, if $\gamma$ is of at least $O(p)$. In this respect, when $\gamma=O(p)$, the ROC converges to a remarkably simple non-trivial limiting profile as $p$ increases. This interesting observation reveals that, when we have the lowest quality noise covariance estimate at our disposal, the number of plausible signal-bearing samples are beneficial provided that the SNR scales at least linearly with $p$.
Be that as it may, the scenario corresponding to the high dimensional setting is a bit more subtle. To be specific, as $m,n,p\to\infty$ such that $m/p\to c_1\in(0,1)$ and $m/n\to c_2\in(0,1)$ with $\text{SNR}=p\gamma $ (i.e., $\text{SNR}=O(p)$), the leading eigenvalue has detection power above a certain $\gamma$ threshold, whereas below that threshold the leading eigenvalue does not have the detection power. The main reason behind this intriguing behavior is the phase transition phenomena. Notwithstanding the above facts, as $m,n,p\to\infty$ such that $m/p\to c_1\in(0,1)$ and $m=n$, the leading eigenvalue retains its detection power, if SNR is of at least $O(p^2)$. Moreover, the corresponding ROC converges to a simple profile.  

This paper is organized as follows. The signal detection problem in colored noise with an unknown noise covariance matrix has been formulated as a binary hypotheses testing problem in Section II. Section III derives the novel c.d.f. expression for the leading eigenvalue, which is the proposed test statistic for the preceding binary hypothesis testing problem, of non-central $F$-matrix with a rank-one non-centrality parameter. The ROC performance of the leading eigenvalue with respect to the system parameters (i.e., $m,n,p$, and SNR) is analyzed in Section IV. Certain important high dimensional statistical characteristics of the ROC are  also derived therein. Finally, conclusive remarks are made in Section V.
 
{\it Notation}: The following notation is used throughout this paper. 
A complex Gaussian random vector $\boldsymbol{y}\in\mathbb{C}^m$ with mean $\boldsymbol{\mu}\in\mathbb{C}^{m}$ and positive definite covariance matrix $\boldsymbol{\Sigma}\in\mathbb{C}^{m\times m}$ is denoted by $\boldsymbol{y}\sim \mathcal{CN}_m(\boldsymbol{\mu},\boldsymbol{\Sigma})$. 
The superscript $(\cdot)^H$ indicates the Hermitian transpose, and $\text{tr}(\cdot)$ represents the trace of a square matrix. If $y_k\sim \mathcal{CN}_m(\boldsymbol{\mu}_k,\boldsymbol{\Sigma}),k=1,2,\ldots,N$, are independent, then $\sum_{k=1}^N \boldsymbol{y}_k\boldsymbol{y}_k^H$ is said to follow a complex non-central Wishart distribution denoted by $\mathcal{CW}_m\left(N,\boldsymbol{\Sigma},\boldsymbol{\Omega}\right)$, where $\boldsymbol{\Omega}=\boldsymbol{\Sigma}^{-1}\boldsymbol{M}\boldsymbol{M}^H$ with $\boldsymbol{M}=\left(\boldsymbol{\mu}_1\;\ldots\;\boldsymbol{\mu}_N\right)\in\mathbb{C}^{m\times N}$ is the non-centrality parameter. The Gaussian $Q$ function is denoted by $\mathcal{Q}(\cdot)$. The real part of a complex number $z$ is represented as $\Re(z)$ and the magnitude of $z$ is given by $|z|$. The operator $e^{\text{tr}(\cdot)}$ is compactly represented as $\text{etr}(\cdot)$. The mathematical expectation operator is represented as $\mathbb{E}\{\cdot\}$, whereas the probability of an event $\mathcal{A}$ is denoted by $\Pr\left\{\mathcal{A}\right\}$. The symbol $\otimes$ is used to represent the Kronecker product between two matrices.
The $m\times m$ identity matrix is represented by $\boldsymbol{I}_m$. The Euclidean norm of a vector $\boldsymbol{a}\in\mathbb{C}^n$ is denotes by $||\boldsymbol{a}||$, whereas the $2-$norm of a matrix $\boldsymbol{A}\in\mathbb{C}^{n\times n}$ (i.e., the leading singular value of $\boldsymbol{A}$) is denoted by $||\boldsymbol{A}||_2$. 
A diagonal matrix with the diagonal entries $a_1,a_2,\ldots, a_n$ is denoted by $\text{diag}(a_1, a_2,\ldots,a_n)$. A Hermitian positive definite matrix $\boldsymbol{A}\in\mathbb{C}$ is denoted by $\boldsymbol{A} \succ \boldsymbol{0}$. For two Hermitian positive definite matrices $\boldsymbol{A}\in\mathbb{C}^{n\times n}$ and $\boldsymbol{B}\in\mathbb{C}^{n\times n}$, the notation $\boldsymbol{A}\preccurlyeq \boldsymbol{B}$ implies that $\boldsymbol{B}-\boldsymbol{A}$ is positive semi-definite and is denoted by $\boldsymbol{B}-\boldsymbol{A} \succcurlyeq \boldsymbol{0} $. The inverse square root of a Hermitian positive definite matrix is denoted by $(\cdot)^{-1/2}$.
The determinant of an $n\times n$ matrix with its  $i,j$th entry given by $a_{i,j}$ is represented by $\det\left[a_{i,j}\right]_{i,j=1,\ldots,n}$, whereas the determinant of a matrix $\boldsymbol{A}\in\mathbb{C}^{n\times n}$ is denoted by $\text{det}\left[\boldsymbol{A}\right]$. Finally, we use the following notation to compactly represent the
determinant of an $n\times n$ block matrix:
\begin{equation*}
\begin{split}
\det\left[a_{i,j}\;\; b_{i}\right]_{\substack{i=1,2,\ldots,n\\
j=1,2,\ldots,n-1}}&=\left|\begin{array}{ccccc}
 a_{1,1} & a_{1,2}& a_{1,3}& \ldots & b_{1}\\
  %a_{2} & b_{2,2}& b_{2,3}& \ldots & b_{2,n}\\
  \vdots & \vdots & \vdots &\ddots & \vdots \\
  a_{n,1} & a_{n,2}& a_{n,3}& \ldots & b_{n}
 \end{array}\right|.
 \end{split}
\end{equation*}

\section{Detection Problem formulation}
 	Consider the general linear signal observation model:
  \begin{align}  
  \label{eq sensing}
\boldsymbol{x}_i=\boldsymbol{a}s_i+\boldsymbol{n}_i,\;\;i=1,2,\ldots,p,
  \end{align}
	where $p$ is the number of independent observations (i.e., samples), $\boldsymbol{x}_i\in\mathbb{C}^{m}$, $\boldsymbol{a}\in\mathbb{C}^{m}$ is an unknown non-random vector, $s_i\in\mathbb{C}$ is the {\it effective} signal which may be known or unknown, and  $\boldsymbol{n}_i\sim \mathcal{CN}_m(\mathbf{0}, \boldsymbol{\Sigma})$ denotes the colored noise in which the noise covariance matrix $\boldsymbol{\Sigma}$ is also unknown to the detector. It turns out that the above observation model encompasses, among others, various practically important sensing technologies. To be specific, for MIMO radar sensing applications (see e.g., \cite{tang2020polyphase,liu2018bayesian,cui2013,li2007mimo,haimovich2007mimo,fishler2006} and references therein) with colocated uniform linear arrays of transmit and receive antennas, the effective transmitted signal $s_i$ admits the form $s_i=\boldsymbol{b}^H \boldsymbol{c}_i$, where $\boldsymbol{b}\in \mathbb{C}^{N_T}$ is the transmit array steering vector of unit magnitude (i.e., $||\boldsymbol{b}||=1$) and $\boldsymbol{c}_i=\left(c_{1i}\; c_{2i}\;\ldots\; c_{N_T i}\right)^T\in\mathbb{C}^{N_T}$ is the {\it known} discrete-time base-band signal transmitted across $N_T$ transmit antennas. 
Moreover, under this setting, $\boldsymbol{a}$ specializes to $\boldsymbol{a}=\varsigma \boldsymbol{a}_r$, where $\boldsymbol{a}_r$ is the receive array steering vector such that $||\boldsymbol{a}_r||=1$ and $\varsigma\in\mathbb{C}$ is an unknown but deterministic target amplitude. As such, (\ref{eq sensing}) specializes to\footnote{Although the transmit and receive beam steering vectors are commonly parameterized as $\boldsymbol{b}(\theta)$ and $\boldsymbol{a}_r(\theta)$ to demonstrate their explicit dependency on the target direction at $\theta$, here we do not adopt that notation.}
\begin{align}
    \label{sensing MIMO}
\boldsymbol{x}^{\text{MIMO Radar}}_i=\varsigma \boldsymbol{a}_r\boldsymbol{b}^H\boldsymbol{c}_i+\boldsymbol{n}_i,\;\;i=1,2,\ldots,p,
\end{align}
where $p$ is interpreted as the information block length and $\boldsymbol{n}_i$ denotes the Gaussian clutter with an unknown covariance matrix. In contrast, for conventional phase-array radar systems, the effective transmitted signal $s_i$ admits the form \cite{fishler2006} $s_i=\boldsymbol{b}^H\boldsymbol{1}c_{1i}$, where $\boldsymbol{1}=\left(1\;1\;\ldots\;1\right)^T\in\mathbb{R}^{N_T}$. This stems from the fact that the same signal is transmitted by all antennas in a conventional phase-array \cite{li2007mimo}. In light of the above development, for conventional phase-array radar sensing applications, (\ref{eq sensing}) specializes to 
\begin{align}
    \label{sensing PhaseArray}
\boldsymbol{x}^{\text{Phase Array}}_i=\varsigma \boldsymbol{a}_r\boldsymbol{b}^H\boldsymbol{1}c_{1i}+\boldsymbol{n}_i,\;\;i=1,2,\ldots,p,
\end{align}
 where $p$ is the number of samples and again, $\boldsymbol{n}_i$ denotes the Gaussian clutter with an unknown covariance matrix \cite{rong2022adaptive}. The above two important sensing applications further highlight the utility of the generic observation model in (\ref{eq sensing}). 

Now to facilitate further analysis, noting that $\boldsymbol{x}_i,\; i=1,2,\ldots,p$, are independent observations, we may represent them in matrix form as
\begin{align}
    \label{matrix observations}
\boldsymbol{X}=\boldsymbol{a}\boldsymbol{s}^T+\boldsymbol{N}
\end{align}
where $\boldsymbol{X}=\left(\boldsymbol{x}_1\; \boldsymbol{x}_2\;\ldots\;\boldsymbol{x}_p\right)\in\mathbb{C}^{m\times p}$, $\boldsymbol{s}=\left(s_1\;s_2\;\ldots\;s_p\right)^T\in\mathbb{C}^p$, and $\boldsymbol{N}=\left(\boldsymbol{n}_1\; \boldsymbol{n}_2\;\ldots\;\boldsymbol{n}_p\right)\in\mathbb{C}^{m\times p}$. Consequently, the signal detection problem can be written as the following binary hypotheses testing problem
\begin{align*}
	&\mathcal{H}_0:\; \boldsymbol{X}=\boldsymbol{N}\;\;\;\;\;\; \text{Signal is absent}\\
	& \mathcal{H}_1:\; \boldsymbol{X}=\boldsymbol{as}^T+\boldsymbol{N} \;\;\;\;\; \text{Signal is present}.
	\end{align*}
Since we have 
\begin{align}
    \mathbb{E}\left\{\boldsymbol{XX}^H\right\}=||\boldsymbol{s}||^2\boldsymbol{a}\boldsymbol{a}^H+p\boldsymbol{\Sigma},
\end{align}
 the above binary hypotheses problem can equivalently be  written as  
 \begin{align*}
	&\mathcal{H}_0:\; \mathbb{E}\left\{\boldsymbol{XX}^H\right\}=p\boldsymbol{\Sigma}\;\;\;\;\;\; \text{Signal is absent}\\
	& \mathcal{H}_1:\; \mathbb{E}\left\{\boldsymbol{XX}^H\right\}=||\boldsymbol{s}||^2\boldsymbol{a}\boldsymbol{a}^H+p\boldsymbol{\Sigma} \;\;\;\;\; \text{Signal is present}.
	\end{align*}
 If the noise covariance matrix $\boldsymbol{\Sigma}$ were known in advance, pre-whitened observations would have resulted in
 \begin{align*}
	&\mathcal{H}_0:\; \boldsymbol{\Sigma}^{-1/2}\mathbb{E}\left\{\boldsymbol{XX}^H\right\}\boldsymbol{\Sigma}^{-1/2}=p\boldsymbol{I}_m\;\;\;\;\;\; \text{Signal is absent}\\
	& \mathcal{H}_1:\; \boldsymbol{\Sigma}^{-1/2}\mathbb{E}\left\{\boldsymbol{XX}^H\right\}\boldsymbol{\Sigma}^{-1/2}=||\boldsymbol{s}||^2 \boldsymbol{\Sigma}^{-1/2}\boldsymbol{a}\boldsymbol{a}^H \boldsymbol{\Sigma}^{-1/2}+p\boldsymbol{I}_m \;\;\;\;\; \text{Signal is present}.
	\end{align*}
 Clearly, the presence of a signal  is characterized by rank-one perturbation of the scaled identity matrix. Therefore,  the eigenvalues of 
 $\boldsymbol{\Sigma}^{-1/2}\mathbb{E}\left\{\boldsymbol{XX}^H\right\}\boldsymbol{\Sigma}^{-1/2}$ are given by $p+||\boldsymbol{s}||^2 \boldsymbol{a}^H\boldsymbol{\Sigma}^{-1}\boldsymbol{a}$ and $p$ (repeated $m-1$ times). This in turn reveals that it is convenient to use the leading eigenvalue of $\boldsymbol{\Sigma}^{-1/2}\mathbb{E}\left\{\boldsymbol{XX}^H\right\}\boldsymbol{\Sigma}^{-1/2}$ to detect the presence of a signal. This is further highlighted by the fact, which is derived based on purely group invariance arguments \cite{johnstoneRoy,muirhead},  that in the absence of detailed knowledge about either the signal or the channel, a generic test depends on the eigenvalues of $\boldsymbol{\Sigma}^{-1/2}\mathbb{E}\left\{\boldsymbol{XX}^H\right\}\boldsymbol{\Sigma}^{-1/2}$. 

In practice, the population average $\mathbb{E}\left\{\boldsymbol{XX}^H\right\}$ and covariance matrix $\boldsymbol{\Sigma}$ are unknown so that the above procedure cannot be trivially applied. To circumvent this difficulty, the corresponding population covariance matrices are commonly replaced by their sample estimates. In particular,  $\mathbb{E}\left\{\boldsymbol{XX}^H\right\}$ is replaced by 
\begin{align}
    \widehat{\boldsymbol{R}}=\frac{1}{p}\boldsymbol{XX}^H=\frac{1}{p}\sum_{k=1}^p \boldsymbol{x}_k\boldsymbol{x}_k^H,
\end{align}
whereas the sample estimate of $\boldsymbol{\Sigma}$ is computed based on the assumption of the availability of the so-called noise-only (i.e., signal-free) additional $n$ training samples $\boldsymbol{z}_1,\boldsymbol{z}_2,\ldots,\boldsymbol{z}_n\in\mathbb{C}^m$, as
\begin{align}
    \widehat{\boldsymbol{\Sigma}}=\frac{1}{n}\sum_{\ell=1}^n \boldsymbol{z}_\ell\boldsymbol{z}_\ell^H,\;\; n\geq m.
\end{align}
This particular assumption about the availability of noise-only or secondary-data samples is commonly used in the literature to facilitate computing the sample estimate of the unknown noise (also clutter) covariance matrix \cite{rong2022adaptive,richmond,zachariah,dogandzic2003generalized,hemimo,chamain2020eigenvalue,nadakuditi}. Moreover, it is noteworthy that the condition $n\geq m$ ensures the almost sure invertibility of the sample covariance matrix $\widehat{\boldsymbol{\Sigma}}$. In this respect, a case of particular interest is when $n=m$, for which $\widehat{\boldsymbol{\Sigma}}$ is a poor estimate of $\boldsymbol{\Sigma}$ but yet invertible. Therefore, the parameter $n-m$ can be considered as an explicit indicator of the quality of the sample covariance estimate. Having estimated the respective sample covariance matrices, now we can conveniently focus on a test based on the leading eigenvalue of  $\widehat{\boldsymbol{\Sigma}}^{-1/2} \widehat{\boldsymbol{R}} \widehat{\boldsymbol{\Sigma}}^{-1/2}$ instead.

The leading sample eigenvalue (a.k.a. Roy's largest root) has been popular among the detection theorists (see e.g., \cite{Kritchman,YCliang,prathapRoyRoot,larson,wang2017stat} and references therein). This is further highlighted, as delineated in \cite{johnstoneRoy}, by the fact that the Roy's largest root test is most powerful among the common tests when the alternative is of rank-one \cite{Kritchman}. Therefore, in light of the above discussion, we choose
$\lambda_{\max}\left(\widehat{\boldsymbol{\Sigma}}^{-1/2}\boldsymbol{\widehat R}\widehat{\boldsymbol{\Sigma}}^{-1/2}\right)=\lambda_{\max}\left(\widehat{\boldsymbol{\Sigma}}^{-1}\boldsymbol{\widehat R}\right)$
as the test statistic\footnote{Although the two matrices $\widehat{\boldsymbol{\Sigma}}^{-1/2}\boldsymbol{\widehat R}\widehat{\boldsymbol{\Sigma}}^{-1/2}$ and $\widehat{\boldsymbol{\Sigma}}^{-1}\boldsymbol{\widehat R}$ are different in principle, they have the same non-zero eigenvalues.}. Now noting that 
\begin{align*}
	&\mathcal{H}_0:\left\{\begin{array}{cl}
         \boldsymbol{x}_i\sim \mathcal{CN}_m\left(\boldsymbol{0},\boldsymbol{\Sigma}\right)& i=1,2,\ldots,p  \\
        \boldsymbol{z}_\ell\sim \mathcal{CN}_m\left(\boldsymbol{0},\boldsymbol{\Sigma}\right)& \ell=1,2,\ldots,n
    \end{array}\right.\\
	& \mathcal{H}_1:\left\{\begin{array}{ll}
         \boldsymbol{x}_i\sim \mathcal{CN}_m\left(\boldsymbol{a}s_i,\boldsymbol{\Sigma}\right)& i=1,2,\ldots,p  \\
        \boldsymbol{z}_\ell\sim \mathcal{CN}_m\left(\boldsymbol{0},\boldsymbol{\Sigma}\right)& \ell=1,2,\ldots,n,
    \end{array}\right.
\end{align*}
 we obtain the following distributions corresponding to the sample covariance matrices
\begin{align*}
	&\mathcal{H}_0:\left\{\begin{array}{cl}
 p\widehat{\boldsymbol{R}}\sim \mathcal{CW}_m\left(p,\boldsymbol{\Sigma},\boldsymbol{0}\right)\\
         n\widehat{\boldsymbol{\Sigma}}\sim \mathcal{CW}_m\left(n,\boldsymbol{\Sigma}, \boldsymbol{0}\right)
    \end{array}\right.\\
	& \mathcal{H}_1:\left\{\begin{array}{ll}
         p\widehat{\boldsymbol{R}}\sim \mathcal{CW}_m\left(p,\boldsymbol{\Sigma},\boldsymbol{\Omega}\right)  \\
        n\widehat{\boldsymbol{\Sigma}}\sim \mathcal{CW}_m\left(n,\boldsymbol{\Sigma}, \boldsymbol{0}\right)
    \end{array}\right.
\end{align*}
where $\boldsymbol{\Omega}=||\boldsymbol{s}||^2\boldsymbol{\Sigma}^{-1}\boldsymbol{a}\boldsymbol{a}^H$ is the rank-one non-centrality parameter matrix. Under the above Gaussian setting,  the matrix $\widehat{\boldsymbol{\Sigma}}^{-1}\boldsymbol{\widehat R}$ is referred to as the $F$-matrix \cite{johnstoneRoy,wang2017stat,muirhead}. In particular, under $\mathcal{H}_0$ it is called a central $F$-matrix, whereas under $\mathcal{H}_1$ it becomes a non-central $F$-matrix. Moreover, for $p<m$, they are referred to as singular central $F$-matrix and singular non-central $F$-matrix, respectively. These $F$-matrices are instrumental in various statistical decision theoretic applications including one-way multivariate analysis of variance (MANOVA), testing linear restrictions on the multivariate linear model, and canonical correlation analysis (CCA), see e.g.,  \cite{johnstoneRoy,johnstone2020,muirhead,mardia2024,hou2023spiked,asendorf2017,Guan} and references therein.  

Let us, for notational concision, denote the maximum eigenvalue $\lambda_{\max}\left(\widehat{\boldsymbol{\Sigma}}^{-1}\boldsymbol{\widehat R}\right)$ as $\hat{\lambda}_{\max}$. Therefore, the  test based on the leading eigenvalue detects a signal if $\hat{\lambda}_{\max}>\xi_{\text{th}}$, where $\xi_{\text{th}}$ is the threshold corresponding to a desired false alarm rate $\alpha\in(0,1)$ given by
\begin{align} 
\label{pf}
\alpha=P_F(\xi_{\text{th}})=\Pr\left(\hat{\lambda}_{\max}>\xi_{\text{th}}|\mathcal{H}_0\right). 
\end{align}
Consequently, the probability of detection admits
\begin{align}
\label{pd}
    P_D(\boldsymbol{\Omega}, \xi_{\text{th}})=\Pr\left(\hat{\lambda}_{\max}>\xi_{\text{th}}|\mathcal{H}_1\right). 
\end{align}
Subsequent elimination of $\xi_{\text{th}}$ yields an explicit functional relationship between $P_D$ and $P_F$, which is referred to as the ROC profile. This in turn helps evaluate the performance of the leading eigenvalue based test.

The above computational machinery relies on the availability of the c.d.f.s of $\hat{\lambda}_{\max}$ under both hypotheses. To this end, we need to statistically characterize the c.d.f. of the leading eigenvalue of  $\widehat{\boldsymbol{\Sigma}}^{-1}\boldsymbol{\widehat R}$ which is the focus of the following section.

\section{C.D.F. of the Maximum Eigenvalue}
Here we statistically characterize the leading eigenvalue of a complex non-central $F$-matrix having rank-one underlying non-centrality parameter. In particular, we derive a closed-form c.d.f. expression for the leading eigenvalue. To this end, we require certain fundamental results pertaining to the finite dimensional representation of the joint eigenvalue density of non-central $F$-matrix and Jacobi polynomials which are given in the following subsection.   
\subsection{Preliminaries}
\begin{de}
    Let $\boldsymbol{X}\in\mathbb{C}^{m\times n}$ ($m\leq n$) be distributed as $\boldsymbol{X}\sim\mathcal{CN}_{m,n}
\left(\boldsymbol{M}, \boldsymbol{\Sigma}\otimes \boldsymbol{I}_n\right)$, where $\boldsymbol{\Sigma}\in\mathbb{C}^{m\times m}$ is Hermitian positive definite and $\boldsymbol{M}\in\mathbb{C}^{m\times n}$. Then the matrix $\boldsymbol{W}=\boldsymbol{XX}^H$ is said to follow a complex correlated non-central Wishart distribution \cite{james} $\boldsymbol{W}\sim\mathcal{CW}_m\left(n,\boldsymbol{\Sigma},\boldsymbol{\Theta}\right)$ with the non-centrality parameter $\boldsymbol{\Theta}=\boldsymbol{\Sigma}^{-1}\boldsymbol{MM}^H$. The correlated complex central Wishart distribution, corresponding to $\boldsymbol{M}=\boldsymbol{0}$ (i.e., $\boldsymbol{\Theta}=\boldsymbol{0}$), is denoted by $\boldsymbol{W}\sim\mathcal{CW}_m\left(n,\boldsymbol{\Sigma}\right)$.
\end{de}

\begin{de}\label{hyperdef}
    Let $\boldsymbol{A}\in\mathbb{C}^{m\times m }$ and $\boldsymbol{B}\in\mathbb{C}^{m\times m}$ be two Hermitian non-negative definite matrices. Then the confluent hypergeometric function of two matrix arguments is defined as \cite{james}
   \begin{align*}
{}_1\widetilde{\mathcal{F}}_1\left(a;b;\boldsymbol{A},\boldsymbol{B}\right)=\sum_{k=0}^\infty \frac{1}{k!} \sum_{\kappa} \frac{[a]_\kappa}{[b]_\kappa}\frac{C_\kappa(\boldsymbol{A})C_\kappa(\boldsymbol{B})}{ C_\kappa(\boldsymbol{I}_m)}
  \end{align*} 
  where $C_\kappa(\cdot)$ is the complex Zonal polynomial which is a symmetric, homogeneous polynomial of degree $k$ in the eigenvalues of the argument matrix\footnote{The exact algebraic definition of the Zonal polynomial is tacitly avoided here, since it is not required in the subsequent analysis. More details of the zonal polynomials can be found in \cite{james,takemura}.}, $\kappa=(k_1,\ldots,k_m)$, with $k_i$'s being non-negative integers, is a partition of $k$ such that $k_1\geq\cdots\geq k_m\geq 0$ and $\sum_{i=1}^mk_i=k$. Also the complex hypergeometric coefficient $[\cdot]_\kappa$ is defined as
  \begin{align}
      [n]_\kappa=\prod_{i=1}^m (n-i+1)_{k_i}
  \end{align}
  where $(a)_k=a(a+1)\cdots(a+k-1)$ with $(a)_0=1$ denotes the Pochhammer symbol. It turns out that, for a positive integer $n$, $(-n)_k$ assumes
\begin{align}
\label{eqpoch}
    (-n)_k=\left\{
    \begin{array}{ll}
    \frac{(-1)^k n!}{(n-k)!} & \text{if $0\leq k\leq n$}\\
    0 & \text{if $k>n$}.
    \end{array}\right.
\end{align}
Moreover, when $\boldsymbol{B}=\boldsymbol{I}_m$, the above confluent hypergeometric function of two matrix arguments degenerates into the confluent hypergeometric function of one matrix argument as follows
\begin{align}
{}_1\widetilde{\mathcal{F}}_1\left(a;b;\boldsymbol{A}\right)={}_1\widetilde{\mathcal{F}}_1\left(a;b;\boldsymbol{A},\boldsymbol{I}_m\right).  
\end{align}

In the special case of {\it rank-one} $\boldsymbol{A}$, following the complex analogue of \cite[Corollary 7.2.4]{muirhead}, it can be shown that the confluent hypergeometric function of one matrix argument further degenerates into the confluent hypergeometric function of the first kind as
\begin{align}
\label{eqhypdegenclassic}
{}_1\widetilde{\mathcal{F}}_1\left(a;b;\boldsymbol{A}\right)= {}_1F_1\left(a;b;\text{tr}(\boldsymbol{A})\right).
\end{align}
\end{de}\begin{rk}
    It is noteworthy that the both confluent hypergeometric functions of matrix arguments can alternatively be represented in terms of determinants of size $m$ \cite{khatri,gross}.
\end{rk}
\begin{thm}\label{thmF}
    If $\boldsymbol{R}\sim\mathcal{CW}_m\left(p,\boldsymbol{\Sigma},\boldsymbol{\Theta}\right)$ and $\boldsymbol{S}\sim\mathcal{CW}_m\left(n,\boldsymbol{\Sigma}\right)$ are independently distributed with $p,n\geq m$, then $\boldsymbol{F}=\boldsymbol{S}^{-1/2}\boldsymbol{R}\boldsymbol{S}^{-1/2}$ follows a complex non-central $F$-distribution with density function \cite{james}
    \begin{align}
g(\boldsymbol{F})=K_d(m,n,p)\text{etr}\left(-\boldsymbol{\Omega}\right)\frac{\det^{p-m}[\boldsymbol{F}]}{\det^{p+n}\left[\boldsymbol{I}_m+\boldsymbol{F}\right]}\;{}_1\widetilde{\mathcal{F}}_1\left(n+p;p;\boldsymbol{\Omega}\left(\boldsymbol{I}_m+\boldsymbol{F}^{-1}\right)^{-1}\right)
    \end{align}
    where $\boldsymbol{\Omega}=\boldsymbol{\Sigma}^{-1/2}\boldsymbol{MM}^H\boldsymbol{\Sigma}^{-1/2}$ denotes the symmetric form of $\boldsymbol{\Theta}$, ${}_1\widetilde{\mathcal{F}}_1(\cdot;\cdot;\cdot)$ denotes the confluent hypergeometric function of one matrix argument and 
    \begin{align*}
        K_d(m,n,p)=\pi^{-\frac{m}{2}(m-1)}\prod_{k=1}^m \frac{(p+n-k)!}{(p-k)!(n-k)!}.
    \end{align*}
\end{thm}

  Now the joint eigenvalue distribution of the $\boldsymbol{F}$ matrix is given by the following theorem.
\begin{thm}
    The joint-density of the ordered eigenvalues $0<\lambda_1<\lambda_2<\ldots<\lambda_m<\infty$ of the non-central $\boldsymbol{F}$ matrix is given by \cite{james}
    \begin{align}
        \label{jointpdf}
  f(\lambda_1,\lambda_2,\ldots,\lambda_m)=
  K(m,n,p)\text{etr}\left(-\boldsymbol{\Omega}\right)\prod_{k=1}^m \frac{\lambda_k^{p-m}}{\left(1+\lambda_k\right)^{p+n}}
\Delta^2_m(\boldsymbol{\lambda})
{}_1\widetilde{\mathcal{F}}_1\left(n+p;p;\boldsymbol{\Omega},\left(\boldsymbol{I}_m+\boldsymbol{\Lambda}^{-1}\right)^{-1}\right)
    \end{align}
\end{thm}
where   $\Delta_m(\boldsymbol{\lambda})=\prod_{1\leq i<j\leq m}\left(\lambda_j-\lambda_i\right)$ is the Vandermonde determinant,
$\boldsymbol{\Lambda}=\text{diag}\left(\lambda_1,\lambda_2,\cdots,\lambda_m\right)$, and
    $K(m,n,p)=\pi^{\frac{m}{2}(m-1)}K_d(m,n,p)/\prod_{k=1}^m(m-k)!$.

The following definition of Jacobi polynomial is also useful in the sequel.
\begin{de}\label{defjac}
    Jacobi polynomials can be defined as \cite[Eq. 8.962]{gradshteyn}
    \begin{align}
        P_n^{(a,b)}(x)&=\frac{(a+1)_n}{n!}\; {}_2F_1\left(-n, n+a+b+1;1+a;\frac{1-x}{2}\right)\nonumber\\
        &=\frac{(a+1)_n}{n!}\sum_{k=0}^{n}\frac{(-n)_k(n+a+b+1)_k}{k!(1+a)_k}\left(\frac{1-x}{2}\right)^k
    \end{align}
    where $a,b>-1$ and ${}_2F_1(\cdot,\cdot;\cdot;\cdot)$ is the Gauss hypergeometric function. 
\end{de}
Moreover, the successive derivatives of $P_n^{(a,b)}(x)$ take form
    \begin{align}
    \label{jacobiDerivative}
        \frac{\rm d^k}{{\rm d}x^k}P_n^{(a,b)}(x)=2^{-k}(n+a+b+1)_k P_{n-k}^{(a+k,b+k)}(x).
    \end{align}
The following definition of confluent hypergeometric function of the first kind is instrumental in our subsequent analysis.
\begin{de}
\label{defconfcont}
    The confluent hypergeometric function of the first kind ${}_1F_1(a;b;x)$ assumes the contour integral representation given by \cite[Eq. 6.11.1.2]{erdelyi}
    \begin{align}
    \label{countdef}
        {}_1 F_1(a;b;x)=\frac{\Gamma(b)\Gamma(a-b+1)}{\Gamma(a)}\frac{1}{2\pi {\rm j}}\oint_0^{(1+)}
        e^{x z} z^{a-1}(z-1)^{b-a-1} {\rm d}z,\;\;\;\;\;\; \Re\{a\}>0
    \end{align}
    where the contour is a loop starting (and ending) at $z=0$ and encircling $z=1$ once in the {\it positive} sense, $\rm j=\sqrt{-1}$, and $\Gamma(\cdot)$ is the Gamma function.
\end{de}
Having armed with the above fundamental results, we are now in a position to derive the c.d.f. of the leading eigenvalue of $\boldsymbol{F}$ matrix when the underlying non-centrality parameter is rank one, which is the focus of the following subsection.
\subsection{C.D.F. of the Leading Eigenvalue}
Before proceeding further, it is important to note that, since we are interested in rank-one non-centrality parameter $\boldsymbol{\Theta}$, the matrix $\boldsymbol{\Omega}$ is also Hermitian positive definite rank one. Consequently, we have the decomposition $\boldsymbol{\Omega}=\omega \boldsymbol{uu}^H$, where $\omega=\text{tr}(\boldsymbol{\Omega})=\text{tr}(\boldsymbol{\Theta})>0 $ and $\boldsymbol{u}\in\mathbb{C}^{m}$ is such that $\left|\left|\boldsymbol{u}\right|\right|=1$. Now capitalizing on a counter integral approach due to  \cite{wang2017stat,Mo2012,Forrester2013rmt}, the confluent hypergeometric function of two matrix arguments in (\ref{jointpdf}) can be further simplified\footnote{Alternatively, one can use the contour integral given in Definition 3 to arrive at Corollary 1.} to yield an expression for the joint p.d.f. of the ordered eigenvalues of $\boldsymbol{F}$ matrix as shown in the following corollary. 
\begin{cor}
Let $\boldsymbol{\Omega}=\omega \boldsymbol{uu}^H$ with $\omega>0$ and $||\boldsymbol{u}||=1$. Then the joint p.d.f. of matrix $\boldsymbol{F}$ assumes
\begin{align}
\label{jpdfdegen}
f(\lambda_1,\lambda_2,\ldots,\lambda_m)=
  C(m,n,p)\omega^{1-m}e^{-\omega}&\prod_{k=1}^m \frac{\lambda_k^{p-m}}{\left(1+\lambda_k\right)^{p+n}}
\Delta^2_m(\boldsymbol{\lambda})\nonumber\\
& \times \sum_{k=1}^m\frac{\displaystyle {}_1F_1\left(n+p-m+1;p-m+1;\frac{\omega \lambda_k}{1+\lambda_k}\right)}{\displaystyle \prod_{\substack{j=1\\ j\neq k}}^m\left(\frac{\lambda_k}{1+\lambda_k}-\frac{\lambda_j}{1+\lambda_j}\right)}
\end{align}
where 
\begin{align*}
    C(m,n,p)=K(m,n,p)\frac{(m-1)!(n+p-m)!}{(n+p-1)!(p-m)!}.
\end{align*}
\end{cor}
To facilitate further analysis, noting that the mapping $h:x \mapsto \frac{x}{1+x}$, $x\geq 0$, is order preserving, we may employ the variable transformations
\begin{align*}
    x_k=\frac{\lambda_k}{1+\lambda_k},\;\;\; k=1,2,\ldots,m,
\end{align*}
with the corresponding differentials given by ${\rm d}x_k=(1-x_k)^2 {\rm d}\lambda_k,\;k=1,2,\ldots,m$, in (\ref{jpdfdegen}) with some algebraic manipulations to arrive at the joint density of $0<x_1<x_2<\ldots<x_m<1$ as
\begin{align}
\label{pdftransform}
g(x_1,x_2,\ldots,x_m)=C(m,n,p)\omega^{1-m}e^{-\omega}&\prod_{k=1}^m x_k^\beta (1-x_k)^\alpha \Delta^2_m(\boldsymbol{x})
\sum_{k=1}^m\frac{\displaystyle {}_1F_1\left(p+\alpha+1;\beta+1;\omega x_k\right)}{\displaystyle \prod_{\substack{j=1\\ j\neq k}}^m\left(x_k-x_j\right)}
\end{align}
where, for notational convenience, we have used $\alpha=n-m$ and $\beta=p-m$. One of the key advantages of the above representation is that its amenability to the use of powerful orthogonal polynomial technique in the subsequent c.d.f. analysis. 

We find it convenient to determine the c.d.f. of $x_{\max}$, since the c.d.f. of the leading eigenvalue of $\boldsymbol{F}$ matrix (i.e., $\lambda_{\max}$ or $\lambda_m$) is related to the c.d.f. of $x_{\max}$ (i.e., $x_m$) as
\begin{align}
\label{cdftrans}
    \Pr\left\{\lambda_{\max}\leq t\right\}=\Pr\left\{x_{\max}\leq \frac{t}{1+t}\right\},\;\;\;\;\;\;\;\;\;\;\;\; 0\leq t<\infty.
\end{align}
Therefore, in what follows, we focus on deriving the c.d.f. of $x_{\max}$. To this end, by definition, the c.d.f. of $x_{\max}$ can be written as
\begin{align}
\label{cdfxdef}
    F_{x_{\max}}(t)=\Pr\left\{x_{\max}\leq t\right\}=\idotsint\limits_{0<x_1<x_2\ldots<x_m\leq t} g(x_1,x_2,\ldots,x_m) {\rm d}x_1 {\rm d}x_2\ldots {\rm d}x_m.
\end{align}
The above multiple integral can be evaluated by taking advantage of powerful orthogonal polynomial approach and the contour integral representation given in Definition \ref{defconfcont}  to yield the c.d.f. of $x_{\max}$ and hence the c.d.f. of $\lambda_{\max}$, which is  given by the following theorem.

\begin{thm}\label{thmain}
Let $\boldsymbol{R}\sim\mathcal{CW}_m\left(p,\boldsymbol{\Sigma},\boldsymbol{\Theta}\right)$ and $\boldsymbol{S}\sim\mathcal{CW}_m\left(n,\boldsymbol{\Sigma}\right)$ be independently distributed with {\it rank-one} non-centrality parameter $\boldsymbol{\Theta}$ (also $\boldsymbol{\Omega}$) and $\min(p,n)\geq m$. Then the c.d.f. of the leading eigenvalue $\lambda_{\max}$ of the complex non-central matrix  $\boldsymbol{F}=\boldsymbol{S}^{-1/2}\boldsymbol{R}\boldsymbol{S}^{-1/2}$ is given by
\begin{align}
\label{cdfexact}
    F^{(\alpha)}_{\lambda_{\max}}(t;\omega)=\mathcal{K}(\alpha,\beta,m)
    e^{-\frac{\omega}{1+t}}\left(\frac{t}{1+t}\right)^{m(\alpha+\beta+m)}
    \det\left[\Phi_i^{(\alpha)}(t,\omega) \;\;\;\;\; \Psi_{i,j}(t)\right]_{\substack{i=1,2,\ldots,\alpha+1\\
    j=2,3,\ldots,\alpha+1}}
\end{align}
where $\alpha=n-m$, $\beta=p-m$, $\omega=\text{tr}\left(\boldsymbol{\Omega}\right)=\text{tr}\left(\boldsymbol{\Theta}\right)\geq 0$,
\begin{align*}
 \Phi_i^{(\alpha)}(t,\omega)&=\sum_{k=0}^{\alpha+1-i} \frac{(\alpha+1-i)!(\alpha+\beta+2m+i-2)!}{k!(\alpha+1-i-k)!(\alpha+\beta+2m+i-2-k)!}
    \left(\frac{\omega t}{1+t}\right)^{\alpha-k},\\
    \Psi_{i,j}(t)&=(\beta+m+i-1)_{j-2} P^{(j-2,\beta+j-2)}_{m+i-j}\left(\frac{2}{t}+1\right),
\end{align*}
and 
\begin{align*}
    \mathcal{K}(\alpha,\beta,m)=\prod_{j=0}^{\alpha-1}\frac{(\beta+2m+j-1)!}{(\beta+2m+2j)!}
\end{align*}
with the interpretation  $\mathcal{K}(0,\beta,m)=1$.
\end{thm}
\begin{IEEEproof}
    See Appendix \ref{appa}.
\end{IEEEproof}
It is noteworthy that the computational complexity of the above c.d.f. depends on $\alpha$ through the size of the determinant. Therefore, the above manifestation of the c.d.f. is particularly useful when the difference between $n$ and $m$ is small (i.e., $n-m$ is small), irrespective of their individual  magnitudes. For instance, as shown below, when $n=m$ (i.e., $\alpha=0$), the determinant degenerates into a scalar, thereby giving a concise c.d.f. expression. This is one of the many advantages of using orthogonal polynomial approach. 

An alternative expression for the c.d.f. of $\boldsymbol{F}$ has recently been derived in \cite[Theorem 1]{Dharmawansa2024ISIT}. However, that expression contains a determinant of size $m$, which makes it less numerically efficient for large values of $m$. Moreover, this $m$ size determinantal structure precludes us from identifying remarkably simple degenerative forms corresponding to certain important special configurations (e.g., $m=n$ or $\omega=0$).
In this respect, the following corollaries further highlight the utility of our new $\alpha$ (i.e., $n-m$) dependent determinantal representation.

\begin{cor}\label{corzeroalpha}
    The exact c.d.f. of the leading eigenvalue of $\boldsymbol{F}=\boldsymbol{S}^{-1/2}\boldsymbol{R}\boldsymbol{S}^{-1/2}$ matrix corresponding to $\alpha=0$ (i.e., $m=n$) configuration is given by
    \begin{align}
       F^{(0)}_{\lambda_{\max}}(t;\omega)= e^{-\frac{\omega}{1+t}}\left(\frac{t}{1+t}\right)^{m(\beta+m)}
    \end{align}
\end{cor}
\begin{IEEEproof}
    Proof follows by noting that, for $\alpha=0$, the determinant degenerates into a scalar with $\Phi_{1}^{(0)}(t,\omega)=1$ and $\mathcal{K}(0,\beta,m)=1$. 
\end{IEEEproof}
It turns out that the c.d.f. corresponding to $\alpha=0$ can alternatively be derived purely based on a matrix integral approach as shown in Appendix \ref{appb}.  

The above Corollary \ref{corzeroalpha} turns out to have far reaching ramifications with respect to certain high dimensional characterizations of the leading eigenvalue. To be precise, as $m,n,p\to\infty$ such that $m=n$ and $m/p\to c_1\in(0,1]$, capitalizing on Corollary \ref{corzeroalpha}, we can establish the following stochastic convergence result
    \begin{align}
\lim_{\substack{m,p\to\infty\\ m/p\to c_1\in(0,1]}}F_{\frac{\lambda_{\max}}{m^2}}(x;\omega)=\left\{
\renewcommand{\arraystretch}{1.5}
\begin{array}{ll}
    \exp\left(\displaystyle-\frac{1}{c_1x}\right) & \text{if $\omega/p\to \tau \geq 0$}  \\
     \exp\left(\displaystyle-\frac{\varphi+c_1}{c_1^2x}\right)& \text{if $\omega/p^2\to\varphi\geq 0$}
\end{array}
\right.
    \end{align}
from which we observe that if $\omega=O(p^2)$, then the scaled leading eigenvalue carries some information about the rank-one non-centrality parameter in this particular asymptotic regime. This is in sharp contrast to the observation corresponding to $\omega=O(p)$ scenario. To be specific, for $\omega=O(p)$, the scaled leading eigenvalue cannot discriminate between $\omega=0$ and $\omega>0$. The latter observation is consistent with the result that the phase transition threshold diverges as $m=n$ \cite{Dharmawansa2014} thereby the leading eigenvalue loosing its detection power. Nevertheless, under the scaling $\omega=O(p^2)$, the above new result reveals that the scaled leading eigenvalue retains its discrimination power in the same asymptotic regime.

Another degenerated scenario of Theorem \ref{thmain} of our interest is the case corresponding to $\omega=0$, the c.d.f. of which is given by the following corollary.

\begin{cor}\label{corzeromean}
    The exact c.d.f. of the leading eigenvalue of $\boldsymbol{F}=\boldsymbol{S}^{-1/2}\boldsymbol{R}\boldsymbol{S}^{-1/2}$ matrix corresponding to $\omega=0$ (i.e., central $F$-matrix) configuration is given by
    \begin{align}
       F^{(\alpha)}_{\lambda_{\max}}(t;0)= \mathcal{C}(\alpha,\beta,m)
    \left(\frac{t}{1+t}\right)^{m(\alpha+\beta+m)}
    \det\left[\Psi_{i+1,j+1}(t)\right]_{i,j=1,2,\ldots,\alpha}
    \end{align}
    where 
    \begin{align*}
        \mathcal{C}(\alpha,\beta,m)=\frac{(\alpha+\beta+2m-1)!}{(\beta+2m-1)!}\mathcal{K}(\alpha,\beta,m).
    \end{align*}
\end{cor}
\begin{IEEEproof}
    As per (\ref{cdfexact}), the direct substitution of $\omega=0$ yields all the entries of the first column of the determinant zero except the first entry, which evaluates to $\Phi^{(\alpha)}_1(t,0)=\frac{(\alpha+\beta+2m-1)!}{(\beta+2m-1)!}$. Consequently, we expand the determinant with its first column and shift the indices from $i,j=2,3,\ldots,\alpha+1,$ to $i,j=1,2,\ldots,\alpha,$ which concludes the proof. 
\end{IEEEproof}
It is noteworthy that the above formula for the c.d.f coincides with the previously derived result \cite[Corollary 8]{chamain2020eigenvalue} corresponding to the leading eigenvalue of a central $F$-matrix.

\begin{figure}[t!]
    \centering
    % Subfigure 1
    \begin{subfigure}[b]{0.48\textwidth}
        \centering
        \includegraphics[width=\textwidth]{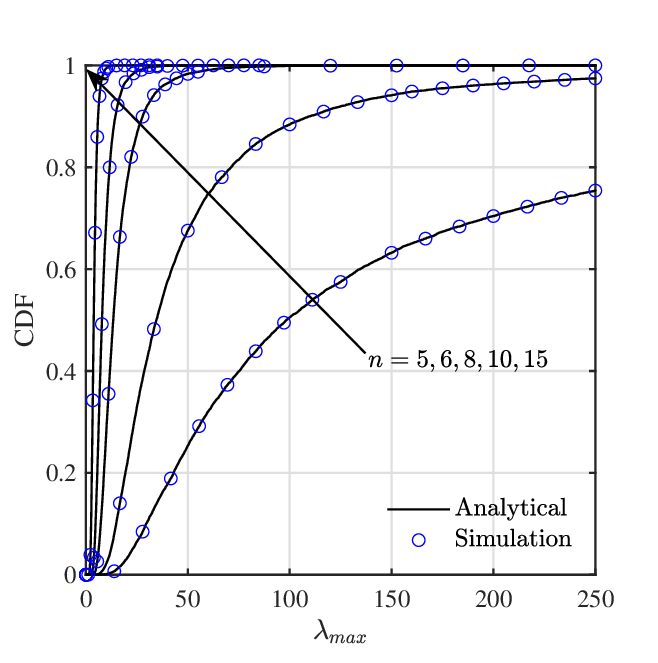}
        \caption{The effect of $n$ for $p=10$.}
        \label{fig1:sub1}
    \end{subfigure}
    \hfill
    % Subfigure 2
    \begin{subfigure}[b]{0.48\textwidth}
        \centering
        \includegraphics[width=\textwidth]{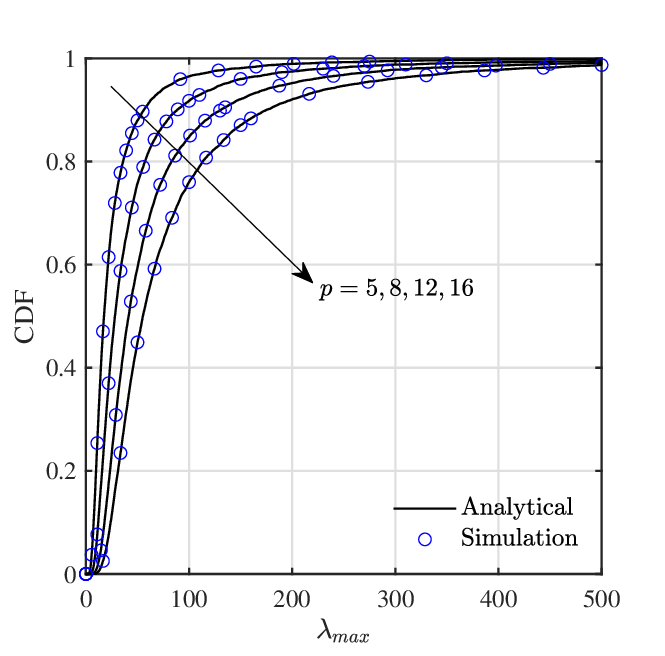}
        \caption{The effect of $p$ for $n=8$.}
        \label{fig1:sub2}
    \end{subfigure}
    
    \caption{Comparison between the theoretical c.d.f. in Theorem \ref{thmain} with simulated values for various system configurations with $m=5$ and $\omega=2$. }
    \label{fig1}
\end{figure}

\begin{figure}[t!]
    \centering
    % Subfigure 1
        \includegraphics[width=0.48\textwidth]{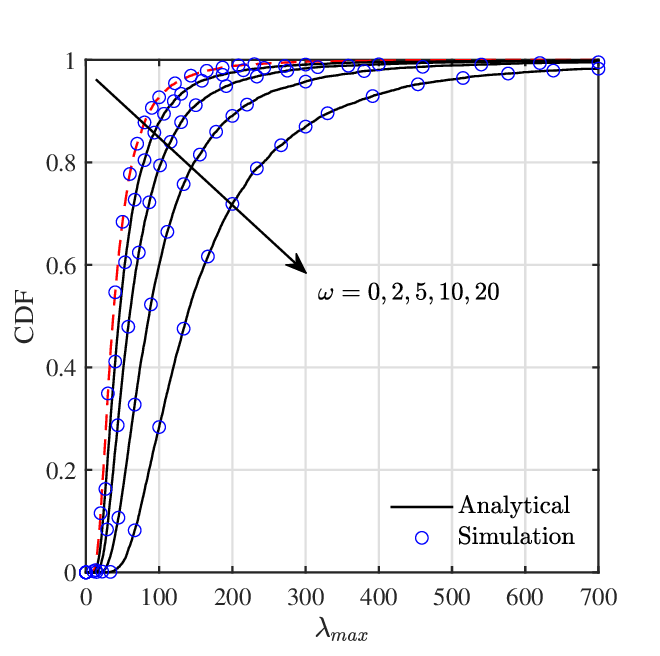}
        \caption{The effect of $\omega$ on the CDF for $m=10, n=12$, and $p=15$. The red dashed curve corresponds to Corollary \ref{corzeromean}.}
        \label{fig2}
\end{figure}

\begin{figure}[t!]
    % Subfigure 1
    \begin{subfigure}[b]{0.48\textwidth}
        \centering
        \includegraphics[width=\textwidth]{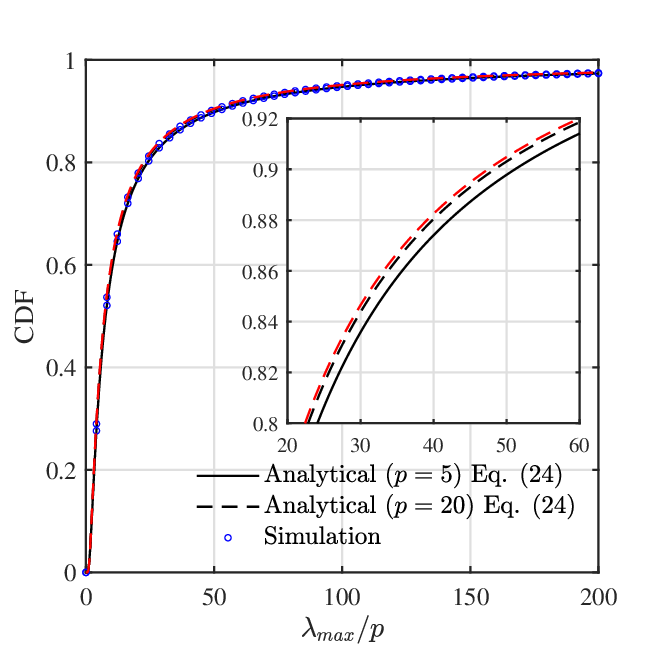}
        \caption{CDF of $\lambda_{\max}/p$ for $m=n=5$ and $\omega=2$. The red dashed curve is the limiting c.d.f. given by $\lim_{p\to\infty}F_{\lambda_{\max}/p}^{(0)}(t;\omega)=e^{-6/t}$.}
        \label{fig3:sub1}
    \end{subfigure}
    \hfill
    \centering
    % Subfigure 1
    \begin{subfigure}[b]{0.48\textwidth}
        \centering
        \includegraphics[width=\textwidth]{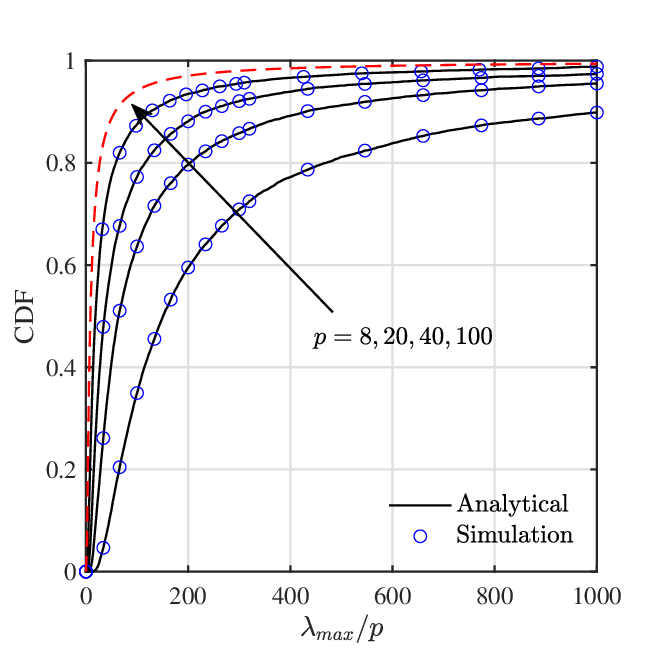}
        \caption{CDF of $\lambda_{\max}/p$ for $m=n=5$ and $\omega=p$. The red dashed curve is the limiting c.d.f. given by $\lim_{p\to\infty}F_{\lambda_{\max}/p}^{(0)}(t;p)=e^{-5/t}$.}
        \label{fig3:sub2}
    \end{subfigure}
    % Subfigure 2

    \caption{The effect of $p$ on the c.d.f. of scaled random variable $\lambda_{\max}/p$ corresponding to the configuration $m=n$ with $\omega=O(1)$ and $\omega=O(p)$.}
    \label{fig3}
\end{figure}

\begin{figure}[t!]
    % Subfigure 1
    \begin{subfigure}[b]{0.48\textwidth}
        \centering
        \includegraphics[width=\textwidth]{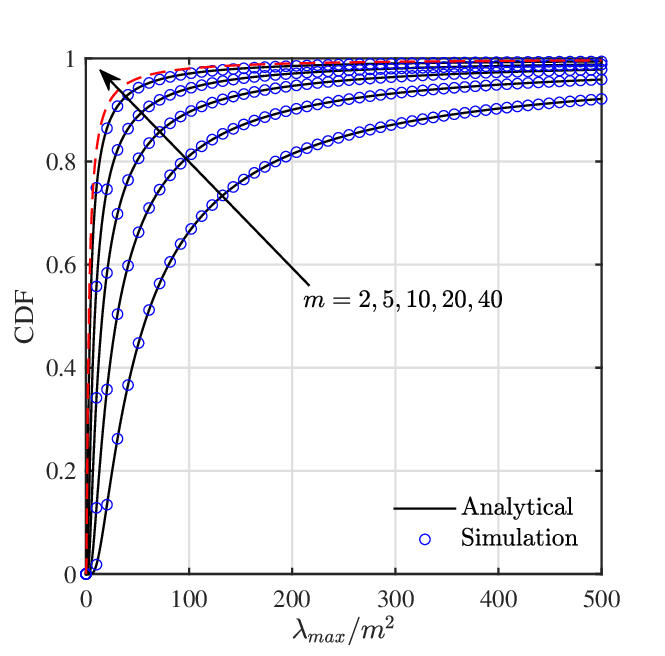}
        \caption{CDF of $\lambda_{\max}/m^2$ for $\omega=p$. The red dashed curve depicts the limiting c.d.f. given by $\lim_{\substack{m,n,p\to\infty\\ m/p\to 1/2}}F_{\lambda_{\max}/p}^{(0)}(t;p)=e^{-2/t}$}
        \label{fig4:sub1}
    \end{subfigure}    
    \hfill
    % Subfigure 2
    \begin{subfigure}[b]{0.48\textwidth}
        \centering
        \includegraphics[width=\textwidth]{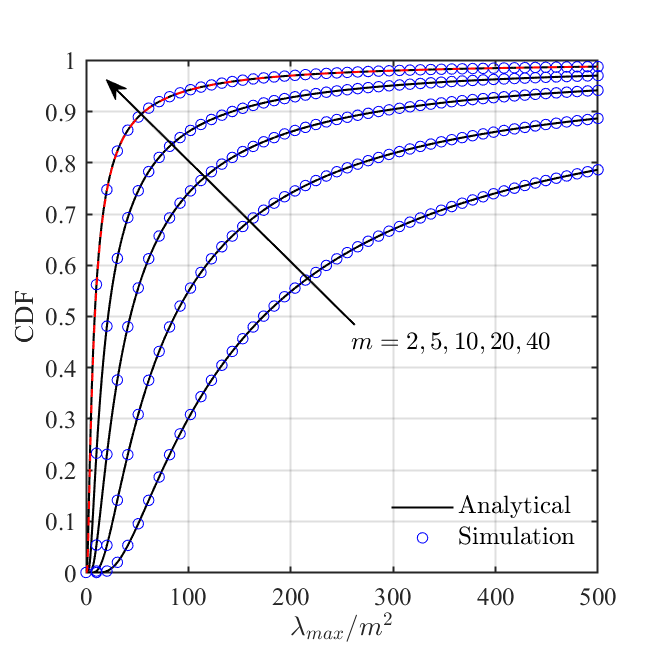}
        \caption{CDF of $\lambda_{\max}/m^2$ for $\omega=p^2$. The red dashed curve depicts the limiting c.d.f. given by $\lim_{\substack{m,n,p\to\infty\\ m/p\to 1/2}}F_{\lambda_{\max}/p}^{(0)}(t;p^2)=e^{-6/t}$}
        \label{fig4:sub2}
    \end{subfigure}

    \caption{Limiting distributions of the scaled maximum eigenvalue $\lambda_{\max}/m^2$ as $m,n,p\to\infty $ such that $m=n$ and $m/p\to 1/2$ with $\omega=O(p)$ and $\omega=O(p^2)$.}
    \label{fig4}
\end{figure}

Figure \ref{fig1} compares the analytical c.d.f. expression given by Theorem \ref{thmain} with simulated data points for various system configurations under the condition that $\omega=2$. In particular, 
the effect of  $n$ and $p$ on the c.d.f. for $m=5$ is shown therein. The degradation of $\lambda_{\max}$ with increasing $n$ is due to the fact that, for fixed $m$ and $p$, as $n\to\infty$, $\left(\boldsymbol{S}/n\right)^{-1/2}\left(\boldsymbol{R}/n\right)\left(\boldsymbol{S}/n\right)^{-1/2}$ tends to zero almost surely. The effect of $\omega$ on the c.d.f. has been depicted in Fig. \ref{fig2}. Figure \ref{fig3} shows the effect of $p$ on the c.d.f. of the scaled random variable $\lambda_{\max}/p$ for $\omega=O(1)$ and $\omega=O(p)$ scenarios with $m=n$. As can be seen from the figure, as $p$ increases, the c.d.f.s converges to their corresponding limiting c.d.f.s. Although, in principle, we need to take the limit as $p\to \infty$ to obtain the respective limiting c.d.f.s, our numerical results demonstrate that these limits still serve as good approximations for moderately large values of $p$. It is also noteworthy that when $\omega=O(1)$, the limitings c.d.f.s of the scaled $\lambda_{\max}$  converges to the same limit under the both hypotheses. However, when $\omega=O(p)$, as can be seen from the figure, the null and the alternative give rise to two different limiting c.d.f.s. Finally, Fig. \ref{fig4} shows the behavior of the scaled random variable $\lambda_{\max}/m^2$ corresponding to the configurations $m=n$ with $\omega=O(p)$ and $\omega=O(p^2)$ as $m,n,p\to\infty$ such that $m/p\to 1/2$.

Be that as it may, to further demonstrate the utility of our main Theorem \ref{thmain}, we focus on evaluating the following useful matrix integral
\begin{align}
    \mathcal{I}_{\alpha,\beta,\nu}(z)=\int_{\boldsymbol{0}}^{\boldsymbol{I}_m} \text{det}^\beta [\boldsymbol{Y}]{\det}^{\alpha}\left[\boldsymbol{I}_m-z\boldsymbol{Y}\right]\text{tr}^\nu \left(\boldsymbol{AY}\right) {\rm d}\boldsymbol{Y},\;\;\ z\in[0,1]
\end{align}
where $\alpha, \beta,\nu=0,1,2,\ldots$,  $\boldsymbol{A}\in\mathbb{C}^{m\times m}$ is a Hermitian positive semi-definite rank-one matrix such that $\text{tr}(\boldsymbol{A})=a>0$, and $\boldsymbol{Y}\in\mathbb{C}^{m\times m}$ is a Hermitian positive definite matrix. The above matrix integral assumes closed-form solutions for the extreme values of $z$ (i.e., $z=0$ and $z=1$)\cite{james,Mathai,kan,muirhead}. Moreover, for $\nu=0$, it takes form of Gauss hypergeometric function of one matrix argument \cite{james,Mathai,kan,muirhead,dumit,koev}. Since the value of $\mathcal{I}_{\alpha,\beta,\nu}(z)$ for a general parameter configuration of $\alpha, \beta, \nu$ and $z\in(0,1)$ is not available in the literature, we present it in the following proposition.
\begin{pro}\label{pmatint}
    Let $\boldsymbol{A}\in\mathbb{C}^{m\times m}$ be Hermitian positive semi-definite with unit rank and $\text{tr}\left(\boldsymbol{A}\right)=a>0$. Then, for $\alpha,\beta,\nu=0,1,2,\ldots$, and $z\in(0,1)$, we have
    \begin{align}
\int_{\boldsymbol{0}}^{\boldsymbol{I}_m} \text{det}^\beta [\boldsymbol{Y}]{\det}^{\alpha}\left[\boldsymbol{I}_m-z\boldsymbol{Y}\right]\text{tr}^\nu \left(\boldsymbol{AY}\right) {\rm d}\boldsymbol{Y}&=\frac{\mathcal{K}(\alpha,\beta,m)(m+\beta)_\nu}{\mathcal{K}_d(\alpha,\beta,m)(\alpha+\beta+2m)_\nu}a^\nu z^{m\alpha}\nonumber\\
&\qquad  \times \det\left[\zeta_i\;\;\;\;\; \Psi_{i,j}\left(\frac{z}{1-z}\right)\right]
_{\substack{i=1,2,\ldots,\alpha+1\\
j=2,3,\ldots,\alpha+1}}
    \end{align}
    where
    \begin{align}
\zeta_i=\sum_{\ell=0}^{\min\left(\alpha,\nu\right)}(-1)^i
\frac{\nu!(-\ell-1)_i(\alpha+1-i)!(\alpha+\beta+2m+i-2)!}
{(\nu-\ell)!(\ell+1)! (\alpha-\ell)!(\ell+\beta+2m+i-2)!}
    \end{align}
    and $\mathcal{K}_d(\alpha,\beta,m)=K_d(m,m+\alpha,m+\beta)$.
\end{pro}
\begin{IEEEproof}
    See Appendix \ref{appc}.
\end{IEEEproof}

Another important scenario arises when the matrix $\boldsymbol{R}$ is rank deficient (i.e., $p<m$), thereby the matrix $\boldsymbol{F}$ is {\it singular}. Notwithstanding that, for $p<m$, $\boldsymbol{F}$ does not have a density on the space of Hermitian $m\times m$ positive definite matrices (see e.g.,  \cite{uhlig,khatrising,ratnasing,diazsing,mathaising} and references therein), we may utilize the joint density of the non-zero eigenvalues of singular $F$-matrix given by \cite{james} to arrive at the corresponding c.d.f. of $\lambda_{\max}$. In particular,  
    The c.d.f. of the leading eigenvalue of {\it singular} $\boldsymbol{F}$, for rank-one non-centrality parameter, is obtained from $F_{\lambda_{\max}}(t;\omega)$ in Theorem \ref{thmain} by relabelling the parameters as follows
    \begin{align}
        m\to p,\; p\to m, \;\text{and}\;\; n\to n+p-m.
    \end{align}

Having armed with the statistical characteristics of the leading eigenvalue of non-central $F$-matrices, we next focus on the ROC of the leading eigenvalue based detector.

\section{ROC of the Leading Eigenvalue Based Test }
Here we analyze the ROC performance associated with the maximum eigenvalue based test in finite as well as in asymptotic regimes. 
To this end, by exploiting the relationship between the non-zero eigenvalues  of $\widehat{\boldsymbol{\Sigma}}^{-1}\boldsymbol{\widehat R}$ and that of $\boldsymbol{S}^{-1/2}\boldsymbol{R}\boldsymbol{S}^{-1/2}$ given by by $\hat{\lambda}_j=(n/p)\lambda_j$, for $j=1,2,\ldots,m$, we may invoke Theorem \ref{thmain} to express the c.d.f. of the leading eigenvalue of $\widehat{\boldsymbol{\Sigma}}^{-1}\boldsymbol{\widehat R}$ as
\begin{align}
\label{cdfteststat}
    F_{\hat{\lambda}_{\max}}(x)=\Pr\left\{\hat{\lambda}_{\max}\leq x\right\}=F^{(\alpha)}_{\lambda_{\max}}(c x;\omega)
\end{align}
     where $c=p/n$ and $\omega=\text{tr}\left(\boldsymbol{\Omega}\right)=||\boldsymbol{s}||^2 \boldsymbol{a}^H \boldsymbol{\Sigma}^{-1}\boldsymbol{a}$. For the clarity of presentation, we find it convenient to analyze the finite dimensional and asymptotic behaviors of the ROC in two separate sub sections. 
\subsection{Finite Dimensional Analysis}
Here our focus is on the scenario in which the matrix dimensions $m,n$, and $p$ are finite. As such, following Theorem \ref{thmain} and Corollary \ref{corzeromean} along with (\ref{pf}), (\ref{pd}), the false alarm and detection probabilities can be written, respectively, as
\begin{align}
P_F(\xi_{\text{th}})&=1-F^{(\alpha)}_{\lambda_{\max}}(c\xi_{\text{th}};0)\\
P_D(\omega, \xi_{\text{th}})&=1-F^{(\alpha)}_{\lambda_{\max}}(c\xi_{\text{th}};\omega).
\end{align}
Since the c.d.f. under the null is independent of $\boldsymbol{\Sigma}$, we can conveniently 
conclude that the leading eigenvalue test has the constant-false-alarm rate (CFAR) property.
Although the above quantities are important in their own right, obtaining a functional relationship between $P_D$ and $P_F$ (i.e., ROC) by eliminating the dependency on $\xi_{\text{th}}$ seems to be an arduous task. Nevertheless, such an explicit functional relationship exists when $\alpha$ assumes zero as specified in the following corollary.
\begin{cor}
    Let us, for notational concision, represent the false alarm and detection probabilities as $P_F$ and $P_D$ such that their dependency on $\omega$ and $\xi_{\text{th}}$ is tacitly understood. Then, when $\alpha=0$ (i.e., $m=n$), $P_D$ and $P_F$ are functionally related as
    \begin{align}
        \label{eqrocalpha0}
        P_D=1-\left(1-P_F\right)\exp
        \left\{-\omega\left(1-\left[1-P_F\right]^{\frac{1}{m(\beta+m)}}\right)\right\}.
    \end{align}
\end{cor}
It is noteworthy that, since the condition $\alpha=0$ (i.e., the number of noise only samples equals the system dimensionality) narrowly satisfies the positive definiteness of the estimated noise-only covariance matrix, the above ROC represents the worst possible profile among all profiles generated by different $\alpha$ values. Alternatively, (\ref{eqrocalpha0}) corresponds to an achievable lower bound on the ROC profiles generated by various values of $\alpha$.

The above ROC relationship can be used to gain some insights into the effect of $\omega$, thereby $\boldsymbol{\Sigma}$ on $P_D$ for the case corresponding to $\alpha=0$. To this end, noting that  $\boldsymbol{as}^T$ and $m$ are fixed for a given system configuration under the alternative, we may observe that $\omega=\left\vert \left \vert \boldsymbol{s} \right\vert \right\vert^2 \boldsymbol{a}^H\boldsymbol{\Sigma}^{-1}\boldsymbol{a} $ functionally depends on the two variables $\boldsymbol{\Sigma}$ and $p$. Therefore, this amounts to analyzing the effects of $\boldsymbol{\Sigma}$ and $p$ on $P_D$.
To facilitate further analysis, we may rewrite 
\begin{align}
\label{ROCana}
P_D\left(\boldsymbol{\Sigma},p\right)=1-\left(1-P_F\right)\exp
        \left\{-\left\vert \left \vert \boldsymbol{s} \right\vert \right\vert^2 \boldsymbol{a}^H\boldsymbol{\Sigma}^{-1}\boldsymbol{a}\left(1-\left[1-P_F\right]^{\frac{1}{mp}}\right)\right\}.
    \end{align}
Now in light of \cite[Corollary 7.7.4.]{horn}, we may easily obtain, for $P_F\in[0,1]$, $p\geq 2$, and $k=1,2,\ldots,m$,
\begin{align}
P_D\left(\boldsymbol{\Sigma}_x,p\right)\geq P_D\left(\boldsymbol{\Sigma},p\right)\geq P_D\left(\boldsymbol{\Sigma}_y,p\right) \iff  \boldsymbol{\Sigma}_x\preccurlyeq \boldsymbol{\Sigma}\preccurlyeq \boldsymbol{\Sigma}_y \implies \lambda_k(\boldsymbol{\Sigma}_x)\leq \lambda_k(\boldsymbol{\Sigma})\leq \lambda_k(\boldsymbol{\Sigma}_x)
\end{align}
where $\boldsymbol{\Sigma}_x, \boldsymbol{\Sigma}_y \succ 0$. Moreover, if $\boldsymbol{\Sigma}=\sigma^2\boldsymbol{I}_m$, then we obtain the following much stronger result
\begin{align}
P_D\left(\boldsymbol{\Sigma}_x,p\right)\geq P_D\left(\sigma^2\boldsymbol{I}_m,p\right)\geq &P_D\left(\boldsymbol{\Sigma}_y,p\right) \nonumber\\
& \quad  \iff  \boldsymbol{\Sigma}_x\preccurlyeq \sigma^2\boldsymbol{I}_m\preccurlyeq \boldsymbol{\Sigma}_y \iff \lambda_k(\boldsymbol{\Sigma}_x)\leq \sigma^2\leq \lambda_k(\boldsymbol{\Sigma}_y)
\end{align}
which can be restated in view of \cite[Chapter 16.F]{marshall} as
\begin{align}
\left[\lambda_m(\boldsymbol{\Sigma}_x)\; \lambda_{m-1}(\boldsymbol{\Sigma}_x)\;\ldots\; \lambda_1(\boldsymbol{\Sigma}_x)\right]^T&<_w  \sigma^2\left[1\; 1\; \ldots\; 1\right]^T
<_w \left[\lambda_m(\boldsymbol{\Sigma}_y)\; \lambda_{m-1}(\boldsymbol{\Sigma}_y)\;\ldots\; \lambda_1(\boldsymbol{\Sigma}_y)\right]^T\nonumber\\
& \iff P_D\left(\boldsymbol{\Sigma}_x,p\right)\geq P_D\left(\sigma^2\boldsymbol{I}_m,p\right)\geq P_D\left(\boldsymbol{\Sigma}_y,p\right)
\end{align}
where the symbol $<_w$ denotes the weak majorization between two vectors.\footnote{Let $\boldsymbol{x}=[x_m\; x_{m-1}\; \ldots\;x_1]^T$  and $ \boldsymbol{y}=[y_m\; y_{m-1}\; \ldots\;y_1]^T$ be two vectors such that $x_m\geq x_{m-1}\geq \ldots\geq x_1$ and $y_m\geq y_{m-1}\geq \ldots\geq y_1$. Then $\boldsymbol{x}$ is said to be {\it weakly majorized} by $\boldsymbol{y}$ if $\sum_{j=1}^k x_j\leq \sum_{j=1}^k y_j$, $k=1,2,\ldots,m$. More specifically, it is denoted by $\boldsymbol{x}<_w\boldsymbol{y}$ \cite{marshall}.} This elegant result reveals that the disparity between the eigenvalues of an arbitrary noise covariance matrix and unity (since $\sigma^2$ is arbitrary, without loss of generality here we assume $\sigma^2=1$) indicates whether the particular ROC has improved with respect to the ROC profile corresponding to the identity noise covariance (i.e., spatially uncorrelated noise). To be specific, negative disparities imply improvement, whereas positive disparities indicate the opposite. Moreover, this result indicates the intuition that the ROC curve corresponding to the white noise (i.e., identity noise covariance) serves as a benchmark curve with respect to which one can quantify the effect of noise covariance on the ROC.     

Now an analysis of the impact of $p$ on the ROC corresponding to $\alpha=0$ is in order. Since the power of the test depends on the rank-one mean departure from zero, intuitively, the ROC profile given by (\ref{ROCana}) should improve with the increasing $p$ if the growth of $||\boldsymbol{s}||^2$ is at least $O(p)$.  
Therefore, to facilitate further analysis, we make use of the representation $||\boldsymbol{s}||^2=k p^\epsilon$, where $0\leq \epsilon<\infty$ and $k>0$, to rewrite (\ref{ROCana}) as
\begin{align}
\label{eq roc alpha0}
    P_D\left(\boldsymbol{\Sigma},p\right)=1-\left(1-P_F\right)\exp
        \left\{-k\boldsymbol{a}^H\boldsymbol{\Sigma}^{-1}\boldsymbol{a}\rho\left(p\right)\right\}
\end{align}
where $\rho\left(p\right)=p^\epsilon\left(1-\left[1-P_F\right]^{\frac{1}{mp}}\right)$. Now in order to show that $P_D\left(\boldsymbol{\Sigma},p_1\right)\leq P_D\left(\boldsymbol{\Sigma},p_2\right)$ for $p_1\leq p_2$, one needs to establish the fact 
that $\rho(p_1)\leq \rho(p_2)$. To this end, we consider the continuous function $\rho(z)=z^\epsilon\left(1-\left[1-P_F\right]^{\frac{1}{mz}}\right),\; z\geq 2$. Consequently, the first derivative of $\rho(z)$ with respect to $z$ can be written, after some algebraic manipulation, as
\begin{align}
    \frac{\rm d}{{\rm d} z} \rho(z)=\epsilon z^{\epsilon-1}\left(1-(1-P_F)^{\frac{1}{mz}}\left(1+\ln (1-P_F)^{-\frac{1}{mz\epsilon}}\right)\right)
\end{align}
from which we obtain, noting that $\ln z\leq z-1, \; z>0$, 
\begin{align}
    \frac{\rm d}{{\rm d} z} \rho(z)> \epsilon z^{\epsilon-1}\left(1-(1-P_F)^{\frac{1}{mz}\left(1-\frac{1}{\epsilon}\right)}\right),\;\; z\geq 2.
\end{align}
The above inequality implies that the function $\rho(z)$ is monotonically increasing for $\epsilon\geq 1$. This in turn verifies the fact that, for $\epsilon\geq 1$, $p_1\leq p_2$ implies $P_D\left(\boldsymbol{\Sigma},p_1\right)\leq P_D\left(\boldsymbol{\Sigma},p_2\right)$. Therefore, it can be concluded that, if $||\boldsymbol{s}||^2$ is at least of $O(p)$, then the corresponding power profiles improve with increasing $p$. Moreover, capitalizing on the fact that 
\begin{align}
    \lim_{z\to\infty} \rho(z)=\left\{\begin{array}{cl}
         0& \text{if $0 \leq \epsilon<1$}  \\
        -\frac{1}{m}\ln z & \text{if $\epsilon=1$}\\
        \infty & \text{if $\epsilon >1$},
    \end{array}\right.
\end{align}
as $p\to\infty$, the limiting ROC profile can be written as
\begin{align}
\label{ROCmnlim}
\lim_{p\to\infty} P_D\left(\boldsymbol{\Sigma},p\right)=\left\{\begin{array}{cl}
         P_F& \text{if $0 \leq \epsilon<1$}  \\
        1-(1-P_F)^{1+\frac{k}{m}\boldsymbol{a}^H\boldsymbol{\Sigma}^{-1}\boldsymbol{a}} & \text{if $\epsilon=1$}\\
        1 & \text{if $\epsilon >1$}.
    \end{array}\right.
\end{align}
The above result in turn yields, for $\epsilon=1$, the bounds
\begin{align}
  1-(1-P_F)^{1+\frac{k}{m ||\boldsymbol{\Sigma}||_2}}\leq   \lim_{p\to\infty} P_D\left(\boldsymbol{\Sigma},p\right)\leq 1-(1-P_F)^{1+\frac{k}{m}||\boldsymbol{\Sigma}^{-1}||_2}.
\end{align}

What remains is to determine a practical system configuration which achieves the minimum requirement of $||\boldsymbol{s}||^2=O(p)$.
 It turns out that this particular condition is satisfied by certain standard phase-array radar sensing systems. To be specific, noting that, for standard phase-array systems, $\boldsymbol{s}$ assumes $\boldsymbol{s}^T=\boldsymbol{b}^H \boldsymbol{1} \boldsymbol{c}$, we obtain $||\boldsymbol{s}||^2=|\boldsymbol{b}^H\boldsymbol{1}|^2 \sum_{k=1}^p \left|c_{1k}\right|^2$. Now if we take our liberty to choose $c_{1k}$s from a certain constant-envelope modulation scheme (e.g., MPSK), then we get $\left|c_{1k}\right|=\sqrt{E_s}/|\boldsymbol{b}^H\mathbf{1}|,\;k=1,2,\ldots,p$. Consequently, it yields $||\boldsymbol{s}||^2=p E_s=O(p)$, thereby confirming the practical achievability of the expected growth requirement.   

\begin{figure}[t!]
    \centering
    % Subfigure 1
    \begin{subfigure}[b]{0.48\textwidth}
        \centering
        \includegraphics[width=\textwidth]{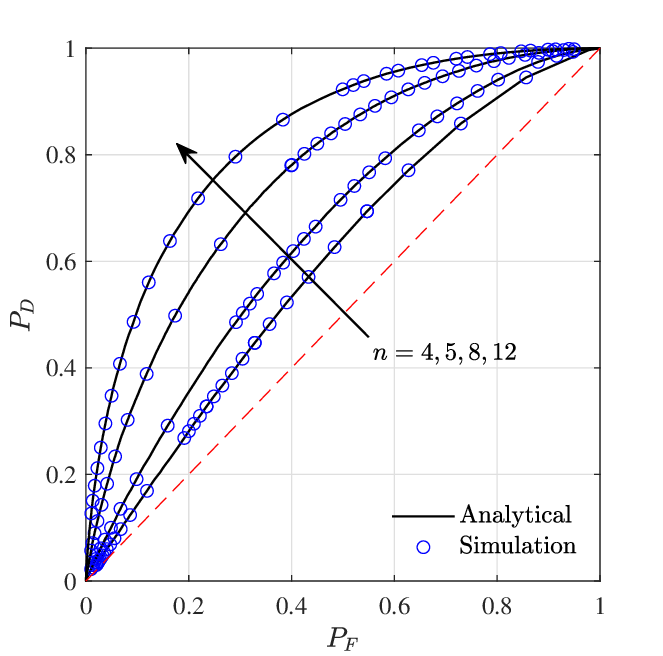}
        \caption{The effect of $n$ for $p=10$ with $\omega=2$.}
        \label{fig5:roc1}
    \end{subfigure}
    \hfill
    % Subfigure 2
    \begin{subfigure}[b]{0.48\textwidth}
        \centering
        \includegraphics[width=\textwidth]{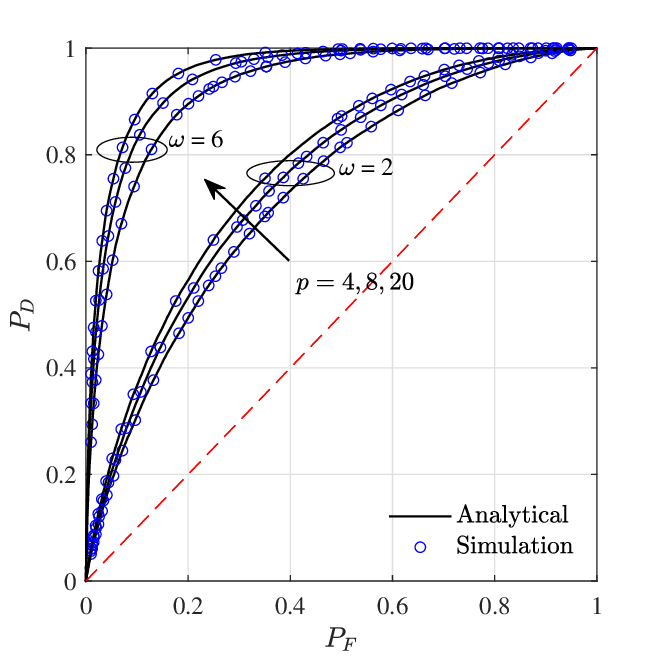}
        \caption{The effect of $p$ and $\omega$ for $n=8$.}
        \label{fig5:roc2}
    \end{subfigure}    
    \caption{The effect of $n,p$ and $\omega$ on ROC profile for $m=4$.}
    \label{fig5}
\end{figure}

\begin{figure}[t!]
    \centering
\includegraphics[width=0.5\textwidth]{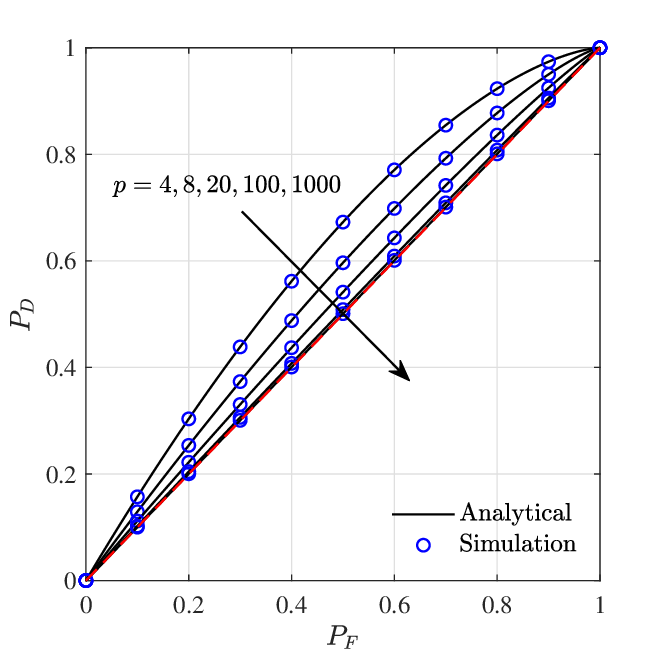}
    \caption{The behavior of ROC profile corresponding to the configuration $m=n=8$ for various values of $p$ with $\omega=2$.}
    \label{fig6}
\end{figure}

\begin{figure}[t!]
    \centering
    % Subfigure 1
    \begin{subfigure}[b]{0.48\textwidth}
        \centering
        \includegraphics[width=\textwidth]{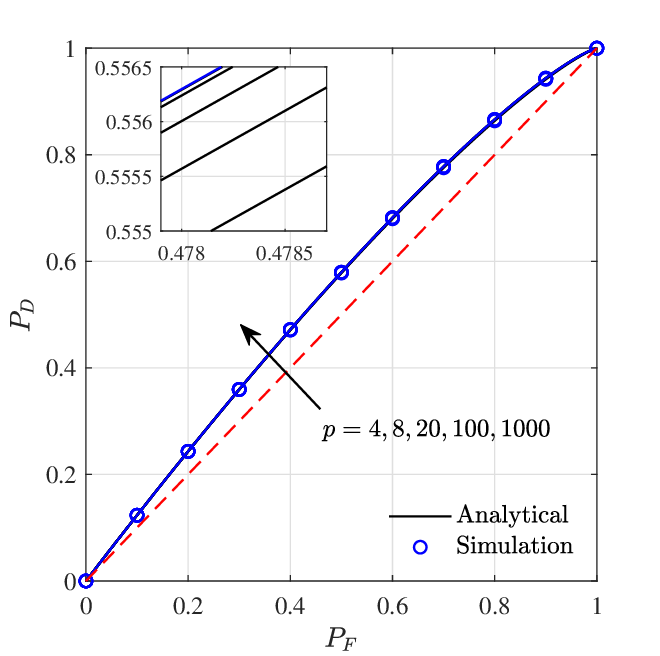}
        \caption{Effect of $p$ for $\omega=p$.}
        \label{fig7:roc_akp}
    \end{subfigure}
    \hfill
    % Subfigure 2
    \begin{subfigure}[b]{0.48\textwidth}
        \centering
        \includegraphics[width=\textwidth]{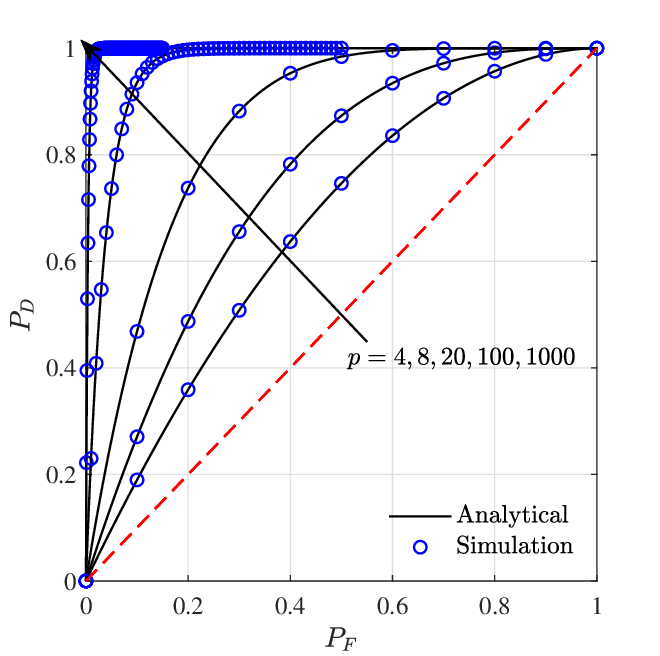}
        \caption{Effect of $p$ for $\omega=p^2$.}
        \label{fig7:roc_akp2}
    \end{subfigure}
    \caption{The effect of $p$ on ROC profiles  for $m=n=4$ with $\omega=O(p)$ and $\omega=O(p^2)$. The blue curve depicts the limiting ROC profile given by $1-(1-P_F)^{5/4}$.}
    \label{fig7}
\end{figure}

Let us now numerically investigate the analytical ROC characteristics derived in the preceding subsection. To be specific,  Fig. \ref{fig5} depicts the effect of $n,p$, and $\omega$, for fixed $m$, on the ROC profiles. As can be seen, for fixed $m$, increase in other parameters improves the ROC profiles. Nevertheless, for the special system configuration of $m=n$ with fixed $\omega$, the ROC profiles tend to degrade with increasing $p$ as shown in Fig. \ref{fig6}. To further investigate this behavior, in Fig. \ref{fig7}, we  present the ROC curves corresponding to two scenarios: $\omega=O(p)$ and $\omega=O(p^2)$. As can be seen from the figure, for $\omega=O(p)$, ROC profiles achieve a limiting curve, which is slightly better than the chance line, as $p$ increases. In contrast, when $\omega=O(p^2)$, ROC profiles tend to the ideal curve as $p$ increases. The above dynamics have been analytically characterized in (\ref{ROCmnlim}).

\subsection{High Dimensional Analysis}
Here we focus on the asymptotic characterization of the ROC of the leading eigenvalue based test.  
In particular, we are interest in the asymptotic regime where $m,n$, and $p$ diverge to infinity such that $m/p\to c_1\in(0,1)$ and $m/n\to c_2\in(0,1)$. Notwithstanding that the c.d.f. expressions we have derived previously are important in the finite dimensional regime, their utility in the above high dimensional regime is severely restricted due to their determinantal structure. To circumvent this difficulty, here we utilize certain tools from the large dimensional random matrix theory of $F$-matrices. In this respect, the high dimensional statistical characteristics of the eigenvalues of central $F$-matrices have been well documented in the literature (see e.g., \cite{nadakuditi,wang2017stat,Johnstone2008,jiang2022invariance,han2016tracy,Guan, jiang2021,WangYao} and references therein). Nevertheless, only a few results are available for the non-central $F$-matrices \cite{Dharmawansa2014,johnstone2020,hou2023spiked,Guan}. As delineated in \cite{Dharmawansa2014}, in the presence of the so-called non-central spikes (i.e., the non-zero eigenvalues of the non-centrality parameter matrix), the eigenvalues of non-central $F$-matrices undergo phase transition, the threshold of which is also derived therein. To be specific, when the population non-central spikes  are below the phase transition threshold (i.e., in the sub-critical regime), then the corresponding sample eigenvalues of a non-central $F$-matrix converges to the right edge of the bulk spectrum and satisfy the most celebrated Tracy-widom law \cite{Guan}, whereas when  when the population non-central spikes are above the phase transition threshold (i.e., in the super-critical regime), then the corresponding sample eigenvalues follow a joint Gaussian density \cite{Dharmawansa2014,hou2023spiked}. This phase transition phenomenon, as we shall show below, 
affects the detection power of the leading eigenvalue based test.

Let us consider a sensing scenario for which $||\boldsymbol{s}||^2=O(p)$ and $||\boldsymbol{\Sigma}^{-1}||_2=O(1)$ as $m,n$, and $p$ diverges to infinity such that $m/p\to c_1\in(0,1)$ and $m/n\to c_2\in(0,1)$. Against this backdrop, the only available non-centrality spike given by  $\omega=||\boldsymbol{s}||^2\boldsymbol{a}^H \boldsymbol{\Sigma}^{-1}\boldsymbol{a}$ takes form $\omega=O(p)$. In this respect, as discussed earlier, for a standard phase-array radar system, $\boldsymbol{s}$ particularizes to $||\boldsymbol{s}||^2=p E_s$. Therefore, in what follows, we investigate the high dimensional characteristics of the leading eigenvalue based test pertaining to the above scenario for which $\omega$ assumes the decomposition $\omega=p\gamma$, where $\gamma=E_s \boldsymbol{a}^H\boldsymbol{\Sigma}^{-1}\boldsymbol{a}$ with $||\boldsymbol{\Sigma}^{-1}||_2=O(1)$.  
Having armed with the above assumptions now we are ready to analyze the high dimensional behavior of the leading eigenvalue based test.

The high dimensional statistical characterizations of $\hat{\lambda}_{\max}$ under the hypotheses $\mathcal{H}_0$ (i.e., signal is absent) and $\mathcal{H}_1$ (i.e., signal is present) are in order now. To this end, capitalizing on large dimensional random matrix theory of central $F$-matrices \cite{nadakuditi,wang2017stat,Johnstone2008,jiang2022invariance,han2016tracy,Guan, jiang2021,WangYao}, as $m,n,p\to\infty$ such that $m/p\to c_1\in (0,1)$ and $m/n\to c_2\in (0,1)$, we have
\begin{align}
\label{tracynull}
    & \mathcal{H}_0:\;\; 
\hat{\lambda}_{\max}\approx \mu + m^{-2/3} \sigma_0 \mathcal{TW}_2
\end{align}
where $\mu=\left(\frac{1+r}{1-c_2}\right)^2$ with $r=\sqrt{c_1+c_2-c_1c_2}$ and 
\begin{align}
    \sigma_0^3=\frac{c_1^4(c_1+r)^4(c_1+c_2)^4}{r \left[(c_1+c_2)^2-c_2(c_1+r)^2\right]^4}
\end{align} are non-random fixed parameters, whereas $\mathcal{TW}_2$ denotes the {\it unitary} Tracy-Widom distributed\footnote{The c.d.f. of $\mathcal{TW}_2$ denoted by $\mathcal{F}_2(t)$ follows the famous Tracy-Widom distribution \cite{tracy1994} corresponding to complex case given by
\begin{align*}
    \mathcal{F}_2(t)=\exp\left(-\int_t^\infty (x-t) q^2(x) {\rm d}x\right)
\end{align*}
in which $q(x)$ denotes the Hastings-McLeod solution of the homogeneous Painlev\'{e} II equation $\frac{{\rm d^2}}{{\rm d} x^2} q(x)=2 q^3(x)+xq(x)$ characterized by the boundary condition $q(x)\sim {\rm Ai}(x)$ as $x\to\infty$, where ${\rm Ai}(x)$ is the Airy function. The Airy function is characterized in turn by $\frac{{\rm d^2}}{{\rm d} x^2} {\rm Ai}(x)=x{\rm Ai}(x)$ and $ {\rm Ai}(+\infty)=0$ \cite{ref:forresterLogGases}.} random variable \cite{tracy1994}. Nevertheless, under the alternative (i.e., $\mathcal{H}_1$), as delineated in \cite{Dharmawansa2014}, phase transition manifests and therefore, $\hat{\lambda}_{\max}$ has two stochastic representations depending on the magnitude of $\gamma$ relative to the phase transition threshold. To be precise, following \cite{Dharmawansa2014,Guan}, we obtain the following stochastic representations
\begin{align}
\label{gaussianfluct}
    \mathcal{H}_1: \hat{\lambda}_{\max}\approx \left\{\begin{array}{ll}\xi + m^{-1/2} \sigma_1 {Z} & \text{if ${\gamma}>\gamma_{p}$}\\
    \mu + m^{-2/3} \sigma_0 {\mathcal {TW}}_2 & \text{if $\gamma<\gamma_{p}$}
    \end{array}\right.
\end{align}
where $Z$ is a standard normal random variable,
\begin{align}
    \sigma_1^2&=\frac{t^2 \gamma^2 (1+\gamma)^2\left(\gamma^2-c_2(1+\gamma)^2-c_1\right) }{\left(c_2-\gamma+c_2 \gamma\right)^4},\\
    \xi&=\frac{(\gamma+c_1)(1+\gamma)}{\gamma-(1+\gamma)c_2},\\
    t^2 &= c_1+c_2-\frac{c_1(\gamma^2-c_1)}{(1+\gamma)^2},
\end{align}
and the phase transition threshold is given by
\begin{align}
    \gamma_{p}=\frac{c_2+r}{1-c_2}.
\end{align}
Here we remark that, since $\xi$ stays away from the upper support of the bulk spectrum\footnote{The limiting spectral density (i.e., bulk spectrum) assumes $\frac{(1-c_2)\sqrt{\left(b-x\right)\left(x-a\right)}}{2\pi x (c_1+c_2x)}$, where $a \leq x\leq b$ with $a=\left(\frac{1-r}{1-c_2}\right)^2$ and $b=\mu=\left(\frac{1+r}{1-c_2}\right)^2$ \cite{Wachter,Silverstein1985}.
} (i.e., $\mu$), we have the strict inequality $\xi>\mu$ \cite{nadakuditi,Dharmawansa2014}.
Now noting that $\gamma=E_s\boldsymbol{a}^H\boldsymbol{\Sigma}^{-1}\boldsymbol{a}$ is equivalent to the SNR, we conveniently refer to $\gamma_p$ as the {\it critical SNR}.
Consequently, we make the key observation that, in the sub-critical-SNR regime (i.e., below the critical SNR $\gamma_p$),  $\hat{\lambda}_{\max}$ has the same asymptotic distribution under the both hypotheses. This in turn reveals that  $\hat{\lambda}_{\max}$ has no detection power asymptotically in the sub-critical-SNR regime. In contrast, in the super-critical-SNR regime (i.e., above the critical SNR $\gamma_p$), the two hypotheses give rise to two different distributions, thereby enabling high precision detection.  

To further verify the above claim related to the detection in the super-critical-SNR regime (i.e., $\gamma>\gamma_p$), for convenience, we consider the following centered and scaled form of $\hat{\lambda}_{\max}$
\begin{align}
\label{test stat1}
t=m^{2/3}\left(\frac{\hat{\lambda}_{\max}-\mu}{\sigma_0}\right).
\end{align}
Now, under the null (i.e., signal free case), as per (\ref{tracynull}), the random variable $t$ follows the unitary Tracy-Widom distribution with the corresponding c.d.f. given by $\mathcal{F}_2(t)$. Therefore, for a given fixed false alarm rate $\alpha\in(0,1)$, we can choose a threshold $t_{\text{th}}$ such that $t_{\text{th}}=\mathcal{F}_2^{-1}(1-\alpha)$. Now what remains is to evaluate the  power of the test in the high-SNR domain. To this end, keeping in mind that, under the alternative (i.e., signal bearing case), for $\gamma>\gamma_p$, $\hat{\lambda}_{\max}$ has Gaussian fluctuations as shown in (\ref{gaussianfluct}), we rearrange the terms in (\ref{test stat1}) to yield
\begin{align}
  \mathcal{H}_1:\; t&\approx m^{2/3}\left(\frac{\xi-\mu+m^{-1/2} \sigma_1 Z}{\sigma_0}\right)= m^{2/3}\left(\frac{\xi-\mu}{\sigma_0}\right)+m^{1/6}\left(\frac{ \sigma_1 Z}{\sigma_0}\right).
\end{align}
Consequently, the asymptotic power of the test can readily be written as
\begin{align}
\label{power1}
    P_D=\Pr\left\{t>t_{\text{th}}\right\}\approx\mathcal{Q}\left(\frac{\sigma_0 t_{\text{th}}-m^{2/3}(\xi-\mu)}{m^{1/6}\sigma_1}\right)
\end{align}
from which we obtain $P_D\to 1$ as $m\to\infty$, since $\xi>\mu$. This further demonstrates that $\hat{\lambda}_{\max}$ based test in (\ref{test stat1}) is asymptotically reliable  in the supercritical SNR regime. Finally, we obtain the following approximate asymptotic ROC profiles
\begin{align}
    P_D\approx \left\{\begin{array}{ll} \displaystyle \mathcal{Q}\left(\frac{\sigma_0 \mathcal{F}_2^{-1}\left(1-P_F\right)-m^{2/3}(\xi-\mu)}{m^{1/6}\sigma_1}\right) & \text{if ${\gamma}>\gamma_{p}$}\\
    P_F & \text{if $\gamma<\gamma_{p}$}.
    \end{array}\right.
\end{align}

Another scenario of theoretical interest is when $m,n,p\to\infty$ such that $m=n$ and $m/p\to c_1\in(0,1)$. Possible consequences in this respect can be understood by studying the behavior of $\hat{\lambda}_{\max}$ as $c_2\to 1$. Clearly, subject to the condition $\omega=O(p)$, the upper support of the bulk (i.e., $\mu=(1+r)^2/(1-c_2)^2$) as well as the phase transition threshold (i.e., $\gamma_p$) diverge to infinity as $c_2\to 1$, thereby $\hat{\lambda}_{\max}$ is lacking the detection power. To further investigate this scenario, let us assume $||\boldsymbol{s}||^2=kp^\epsilon$  with $ 0\leq \epsilon<\infty $ and  $||\boldsymbol{\Sigma}^{-1}||_2=O(1)$ to rewrite (\ref{eq roc alpha0}) as
\begin{align}
    P_D\left(m,p\right)=1-\left(1-P_F\right)\exp
        \left\{-k\boldsymbol{a}^H\boldsymbol{\Sigma}^{-1}\boldsymbol{a}\rho\left(m,p\right)\right\}
\end{align}
where $\rho\left(m,p\right)=p^\epsilon\left(1-\left[1-P_F\right]^{\frac{1}{mp}}\right)$. Now noting the limit
\begin{align}
    \lim_{\substack{m,p\to\infty\\m/p\to c_1\in(0,1)}} \rho(m,p)=\left\{\begin{array}{cl}
         0& \text{if $0 \leq \epsilon<2$}  \\
        -\frac{1}{c_1}\ln \left(1-P_F\right) & \text{if $\epsilon=2$}\\
        \infty & \text{if $\epsilon >2$},
    \end{array}\right.
\end{align}
we finally obtain
\begin{align}
\lim_{\substack{m,p\to\infty\\m/p\to c_1\in(0,1)}}P_D\left(m,p\right)=\left\{\begin{array}{cl}
         P_F& \text{if $0 \leq \epsilon<2$}  \\
        1-(1-P_F)^{1+\frac{k}{c_1}\boldsymbol{a}^H\boldsymbol{\Sigma}^{-1}\boldsymbol{a}} & \text{if $\epsilon=2$}\\
        1 & \text{if $\epsilon >2$}.
    \end{array}\right.
\end{align}
Therefore, for $\epsilon=2$, we the following bounds are in order
\begin{align}
  1-(1-P_F)^{1+\frac{k}{c_1 ||\boldsymbol{\Sigma}||_2}}\leq   \lim_{\substack{m,p\to\infty\\m/p\to c_1\in(0,1)}}P_D\left(m,p\right)\leq 1-(1-P_F)^{1+\frac{k}{c_1}||\boldsymbol{\Sigma}^{-1}||_2}.
\end{align}
The above result demonstrates that, if $\omega=O(p^\epsilon)$ with $\epsilon\geq 2$, then as $m,n,p\to\infty$ such that $m=n$ and $m/p\to c_1\in(0,1)$, the leading eigenvalue retains its detection power.

\begin{figure}[t!]
    \centering
    % Subfigure 1
    \begin{subfigure}[b]{0.48\textwidth}
        \centering
        \includegraphics[width=\textwidth]{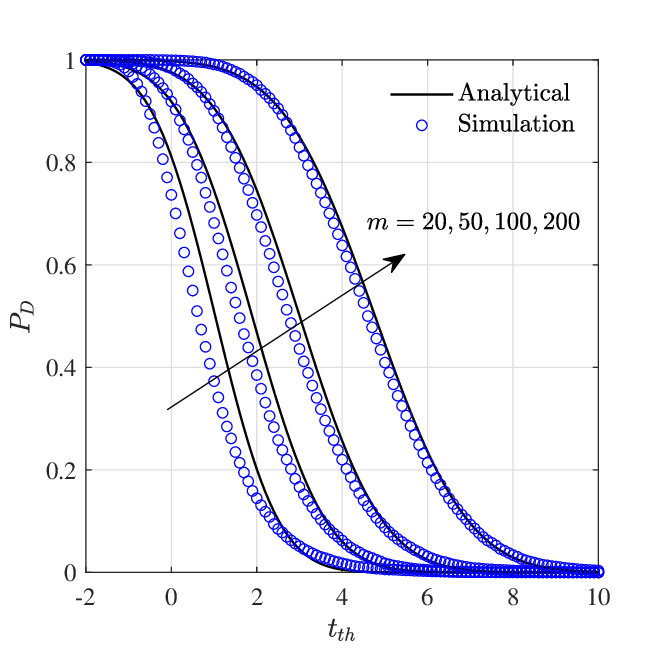}
        \caption{Power of the test.}
        \label{fig8:sub1}
    \end{subfigure}
    \hfill
    % Subfigure 2
    \begin{subfigure}[b]{0.48\textwidth}
        \centering
        \includegraphics[width=\textwidth]{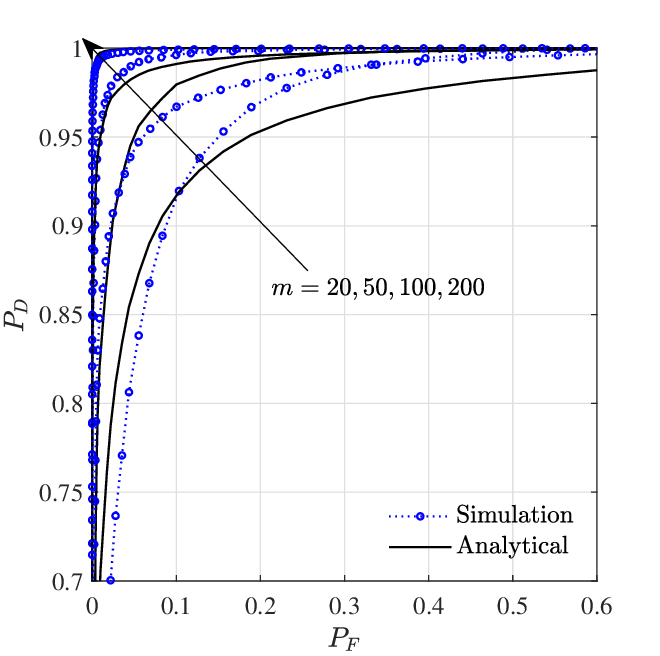}
        \caption{ROC of the test.}
        \label{fig8:sub2}
    \end{subfigure}
    \caption{High dimensional characteristics of the centered and scaled random variable
    $t$ in the supercritical regime (i.e., above the phase transition threshold $\gamma_p$). Results are shown for $\gamma=5>\gamma_p\approx 2.58$ with $c_1=0.25$ and $c_2=0.5$.}
    \label{fig8}
\end{figure}

The high dimensional characteristics of the leading eigenvalue based test are numerically depicted in Fig. \ref{fig8}. In particular, Fig. \ref{fig8:sub1} compares the analytical and simulated results for the power of the test given in (\ref{test stat1}). The high dimensional regime of our interest is above the phase transition threshold (i.e., super-critical regime). The asymptotic Gaussianity of the test in this particular regime is clearly visible in the figure. The ROC profiles corresponding to the same regime are depicted in Fig. \ref{fig8:sub1}. The discrepancies between the asymptotic and simulated ROC results tend to vanish as $m$ increases. It is noteworthy that, despite the fact that those asymptotic analytical results are derived based on the assumption that $m,n,p\to\infty$, our numerical results verify that those analytical results serve as very good approximations in the finite dimensional scenarios as well.

\section{Conclusion}
This paper investigates the signal detection problem in colored noise using the leading eigenvalue of whitened signal-plus-noise sample covariance matrix. To be specific, our focus is on two scenarios: finite dimensional and high dimensional. Corresponding to the finite dimensional scenario, we take advantage of powerful orthogonal polynomial approach in random matrix theory to derive a novel c.d.f. for the leading eigenvalue of non-central $F$-matrix with a rank-one non-centrality parameter. It turns out that the leading eigenvalue based test has the CFAR property. Capitalizing on this new c.d.f. expression we have analyzed the performance of the test by deriving corresponding ROC profiles. Our results reveal that, for fixed $m$ and $n$ such that $m=n$ (i.e., the noise only sample covariance matrix is nearly rank deficient), the power of the test improves with $p$ (i.e., the number of plausible signal-bearing samples) if SNR is of at least $O(p)$. In contrast, when $m,n,p$ diverges such that $m=n$ and $m/p\to c_1\in(0,1)$, the power of the test converges if SNR is of at least $O(p^2)$. However, in the same regime with $m/n\to c_2\in(0,1)$ and $\text{SNR}=O(p)$, the leading eigenvalue exhibits an intriguing behavior; it cannot detect the presence of weak signals. This observation is intimately related to the phase transition phenomena in infinite dimensional random matrix theory. Notwithstanding the above facts, the analysis corresponding to a non-central $F$-matrix with an arbitrary rank non-centrality parameter remains as an open problem.

%%%%%%
%% Appendix:
%% If needed a single appendix is created by
%%
%\appendix
%%

%\appendix

%% If several appendices are needed, then the command
%%
 \appendices
%%
%% in combination with further \section commands can be used.
%%%%%%
\section{Proof of Theorem \ref{thmain}}\label{appa}
Since $g(x_1,x_2,\ldots,x_m)$ is symmetric in $x_1,x_2,\ldots,x_m$, we may remove the ordered region of integration in (\ref{cdfxdef}) to yield
\begin{align}
    F_{x_{\max}}(t)=\frac{1}{m!}\int_0^t\int_0^t\cdots\int_0^t g(x_1,x_2,\ldots,x_m) {\rm d}x_1 {\rm d}x_2\ldots {\rm d}x_m
\end{align}
where 
\begin{align*}
g(x_1,x_2,\ldots,x_m){=}C(m,n,p)\omega^{1-m}e^{-\omega}&\prod_{k=1}^m x_k^\beta (1-x_k)^\alpha \Delta^2_m(\boldsymbol{x})
\sum_{k=1}^m\frac{\displaystyle {}_1F_1\left(p+\alpha+1;\beta+1;\omega x_k\right)}{\displaystyle \prod_{\substack{j=1\\ j\neq k}}^m\left(x_k-x_j\right)}.
\end{align*}
Noting that the above multiple integral consists of a sum of $m$ individual multiple integrals and each such integral evaluates to the same value, we take the liberty of choosing the multiple integral corresponding to the $k=1$ term of the summation to yield
\begin{align}
    F_{x_{\max}}(t)=\frac{C(m,n,p)e^{-\omega}}{(m-1)! \omega^{m-1}}\int_0^t\int_0^t\cdots\int_0^t &\frac{\prod_{k=1}^m x_k^\beta (1-x_k)^\alpha}{ \prod_{j=2}^m\left(x_1-x_j\right)} \Delta^2_m(\boldsymbol{x})\nonumber\\
    & \quad \times
 {}_1F_1\left(p+\alpha+1;\beta+1;\omega x_1\right) {\rm d}x_1 {\rm d}x_2\ldots {\rm d}x_m.
\end{align}
To facilitate further analysis, we may employ the decomposition 
\begin{align}
\Delta^2_m(\boldsymbol{x})=\prod_{j=2}^m(x_1-x_j)^2 \Delta^2_{m-1}(\boldsymbol{x}),
\end{align}
where $\Delta^2_{m-1}(\boldsymbol{x})=\prod_{2\leq i<j\leq m }(x_j-x_i)^2 $, in the above multiple integral with some algebraic manipulation to obtain
\begin{align}
\label{eqmixx}
     F_{x_{\max}}(t)=\frac{C(m,n,p)e^{-\omega}}{(m-1)! \omega^{m-1}}\int_0^t\int_0^t\cdots\int_0^t &x_1^\beta (1-x_1)^\alpha \prod_{k=2}^m x_k^\beta (1-x_k)^\alpha(x_1-x_k) \Delta^2_{m-1}(\boldsymbol{x})\nonumber\\
    & \quad \times
 {}_1F_1\left(p+\alpha+1;\beta+1;\omega x_1\right) {\rm d}x_1 {\rm d}x_2\ldots {\rm d}x_m.
\end{align}
Now in view of making the limits of integration independent of $t$,  we find it convenient to introduce the variable transformations
\begin{align}
    x_1=ty\hspace{2cm} \text{and}\hspace{2cm} x_k=ty_{k-1},\;\;\; k=2,3,\ldots,m,
\end{align}
into (\ref{eqmixx}) with some algebraic manipulation to arrive at
\begin{align}
\label{eqcdfint}
    F_{x_{\max}}(t)=\frac{C(m,n,p)e^{-\omega}}{(m-1)! \omega^{m-1}}t^{m(\beta+m-1)+1}
    \int_0^1 & y^\beta (1-ty)^\alpha {}_1F_1\left(p+\alpha+1;\beta+1;\omega t y\right)
    \mathcal{Q}_{m-1}(t,y) {\rm d}y
\end{align}
where
\begin{align}
    \mathcal{Q}_{m}(t,y)=\int_0^1\int_0^1\ldots\int_0^1
    \prod_{k=1}^m y_k^\beta (1-ty_k)^\alpha(y-y_k) \Delta^2_{m}(\boldsymbol{y}) {\rm d}y_1 {\rm d}y_2\ldots {\rm d}y_m.
\end{align}
Since we are interested in using the orthogonal polynomial approach in random matrix theory to evaluate $\mathcal{Q}_{m}(t,y)$, keeping in mind that Jacobi polynomials $P^{(a,b)}_n(x)$ are orthogonal with respect to the weight $(1-x)^a(1+x)^b$ on the interval $x\in[-1,1]$, we may apply the variable transformations, $z_k=2y_k-1,\; k=1,2,\ldots,m$, into the above multiple integral to yield
\begin{align}
  &\mathcal{Q}_{m}(t,y)\nonumber\\
  &=  \frac{t^{\alpha m}}{2^{\varepsilon_m}}\int_{-1}^1\int_{-1}^1\ldots\int_{-1}^1
    \prod_{k=1}^m (1+z_k)^\beta \left(\frac{2}{t}-1-z_k\right)^\alpha(2y-1-z_k) \Delta^2_{m}(\boldsymbol{z}) {\rm d}z_1 {\rm d}z_2\ldots {\rm d}z_m
\end{align}
where $\varepsilon_m=m(\alpha+\beta+m+1)$.
This multiple integral can be evaluated, as shown in Appendix \ref{appd}, by leveraging the powerful orthogonal polynomial approach advocated in \cite{mehta} to yield
\begin{align}
\label{qm}
    \mathcal{Q}_{m}(t,y)=K_1(\alpha,\beta,m)(1-ty)^{-\alpha}t^{\alpha(m+1)} 
    \det\left[P^{(0,\beta)}_{m+i-1}(2y-1)\;\;\;\;\; \Psi_{i+1,j}\left(\frac{t}{1-t}\right)\right]
_{\substack{i=1,2,\ldots,\alpha+1\\
j=2,3,\ldots,\alpha+1}}
\end{align}
where
\begin{align}
   K_1(\alpha,\beta,m)=\alpha!(\alpha+1)!\prod_{j=1}^{\alpha+1}\frac{(m+j-1)!(m+\beta+j-1)!}{j!(2m+2j+\beta-2)!} \prod_{j=0}^{m-1}
   \frac{j!(j+1)!(\beta+j)!}{(\beta+m+j)!}.
\end{align}
Consequently, keeping in mind that $\mathcal{Q}_{m-1}(t,y)$ is of our interest, we may use (\ref{qm}) in (\ref{eqcdfint}) with  $m$ replaced by $m-1$ to arrive at  
\begin{align}
\label{eqcdfint1}
    F_{x_{\max}}(t)&=\frac{K_2(m,n,p)e^{-\omega}}{(m-1)! \omega^{m-1}}t^{m(\alpha+\beta+m-1)+1}\nonumber\\
    & \qquad \times \int_0^1  y^\beta  {}_1F_1\left(p+\alpha+1;\beta+1;\omega t y\right)\nonumber\\
    &\hspace{3.5cm}\times \det\left[P^{(0,\beta)}_{m+i-2}(2y-1)\;\;\;\;\; \Psi_{i,j}\left(\frac{t}{1-t}\right)\right]
_{\substack{i=1,2,\ldots,\alpha+1\\
j=2,3,\ldots,\alpha+1}}
     {\rm d}y
\end{align}
where $K_2(m,n,p)=K_1(\alpha,\beta,m-1)C(m,n,p)$. Since only the first column of the determinant in the integrand depends on $y$, we may rewrite the above integral as
\begin{align}
\label{eqcdfint2}
    F_{x_{\max}}(t)&=\frac{K_2(m,n,p)e^{-\omega}}{(m-1)! \omega^{m-1}}t^{m(\alpha+\beta+m-1)+1}\det\left[\mathcal{A}_i(\omega t)\;\;\;\;\; \Psi_{i,j}\left(\frac{t}{1-t}\right)\right]
_{\substack{i=1,2,\ldots,\alpha+1\\
j=2,3,\ldots,\alpha+1}}
\end{align}
where 
\begin{align}
    \mathcal{A}_i(\omega t)=\int_0^1  y^\beta  {}_1F_1\left(p+\alpha+1;\beta+1;\omega t y\right)P^{(0,\beta)}_{m+i-2}(2y-1){\rm d}y.
\end{align}
To facilitate further analysis, in view of Definition \ref{defjac}, we may expand the Jacobi polynomial term in the above integrand and perform term-by-term integration with the help of \cite[Eq. 7.613.2]{gradshteyn} to obtain 
\begin{align}
\label{eqA}
   \mathcal{A}_i(\omega t)=\beta! \sum_{k=0}^{m+i-2} \frac{(-m-i+2)_k(p+i-1)_k}{k!(\beta+k+1)!}
{}_1F_1\left(p+\alpha+1;\beta+k+2;\omega t\right).
\end{align}
The direct substitution of (\ref{eqA}) into (\ref{eqcdfint2}) would result in a closed-form expression for the desired c.d.f. Nevertheless, a careful inspection of the resultant expression reveals a pole of order $(m-1)$ at $\omega=0$. Therefore, to circumvent this difficulty by removing the singularity, in what follows, we re-sum the above finite series in (\ref{eqA}).

Let us use Definition \ref{countdef} to rewrite the ${}_1F_1$ function in (\ref{eqA}) as
\begin{align}
{}_1F_1\left(p+\alpha+1;\beta+k+2;\omega t\right)=\frac{(\beta+k+1)!(n-k-1)!}{(p+\alpha)!}\frac{1}{2\pi {\rm j}}\oint_0^{(1+)}
        e^{\omega t z} z^{p+\alpha}(z-1)^{k-n} {\rm d}z
\end{align}
which upon substituting into (\ref{eqA}) with some algebraic manipulation gives
\begin{align}
   \mathcal{A}_i(\omega t)&=\frac{\beta!(n-1)!}{(p+\alpha)!} \sum_{k=0}^{m+i-2} (-1)^k\frac{(-m-i+2)_k(p+i-1)_k}{k!(-n+1)_k} \frac{1}{2\pi {\rm j}}\oint_0^{(1+)}
        e^{\omega t z} z^{p+\alpha}(z-1)^{k-n} {\rm d}z.
\end{align}
It is noteworthy that the ratio $(-m-i+2)_k/(-n+1)_k$ is well defined, since $\max_i (m+i-2)\leq (n-1)$.
Now we interchange the summation and contour integration operators to rewrite
\begin{align}
\label{count1f1}
  \mathcal{A}_i(\omega t)  = \frac{\beta!(n-1)!}{(p+\alpha)!}\frac{1}{2\pi {\rm j}}\oint_0^{(1+)}
        \frac{e^{\omega t z} z^{p+\alpha}}{(z-1)^{n}}G(z)
         {\rm d}z
\end{align}
where
\begin{align}
    G(z)=\sum_{k=0}^{m+i-2}\frac{(-m-i+2)_k(p+i-1)_k}{k!(-n+1)_k} (1-z)^{k}
\end{align}
which in view of \cite[Eq. 2.1.1.4]{erdelyi} assumes
\begin{align}
    G(z)={}_2F_1\left(-m-i+2,p+i-1;-n+1;1-z\right).
\end{align}
In view of further simplifying the contour integral in (\ref{count1f1}), capitalizing on the hypergeometric transformation ${}_2F_1(a,b;c;z)=(1-z)^{c-b-a}{}_2F_1(c-a,c-b;c;z),\; |z|<1$, we obtain
\begin{align}
    G(z)=z^{-p-\alpha}{}_2F_1\left(-\alpha+i-1,-n-p-i+2;-n+1;1-z\right),
\end{align}
which in view of $\alpha-i+1<n+p+i-2$ and $\max_i(\alpha-i+1)<n-1$, can be expanded as 
\begin{align}
\label{eqG}
    G(z)=\frac{z^{-p-\alpha}}{(n-1)!}\sum_{k=0}^{\alpha+1-i}
    (-\alpha+i-1)_k (-n-p-i+2)_k(n-1-k)! \frac{(z-1)^k}{k!}.
\end{align}
Consequently we substitute (\ref{eqG}) into (\ref{count1f1}) and swap the summation and integral operators to yield
\begin{align}
\label{eqcounteval}
  \mathcal{A}_i(\omega t)  &= \frac{\beta!}{(p+\alpha)!}
  \sum_{k=0}^{\alpha+1-i}
    \frac{(-\alpha+i-1)_k (-n-p-i+2)_k}{k!} \frac{(n-1-k)!}{2\pi {\rm j}}\oint_0^{(1+)}
        \frac{e^{\omega t z}}{(z-1)^{n-k}}
         {\rm d}z\nonumber\\
         &=\frac{\beta!e^{\omega t}}{(p+\alpha)!}
  \sum_{k=0}^{\alpha+1-i}
    (-\alpha+i-1)_k (-n-p-i+2)_k \frac{\left(\omega t\right)^{n-k-1}}{k!}
\end{align}
in which the second equality is due to the Cauchy residue theorem. Finally, we substitute (\ref{eqcounteval}) into (\ref{eqcdfint2}) and make use of (\ref{cdftrans}) with some algebraic manipulation to conclude the proof.

\section{An Alternative Proof of Corollary \ref{corzeroalpha} } \label{appb}
The density of $\boldsymbol{F}$, for $n=m$ (i.e., $\alpha=0$), particularizes to 
\begin{align}
g(\boldsymbol{F})=K_d(m,m,p)\text{etr}\left(-\boldsymbol{\Omega}\right)\frac{\det^{p-m}[\boldsymbol{F}]}{\det^{p+m}\left[\boldsymbol{I}_m+\boldsymbol{F}\right]}\;{}_1\widetilde{\mathcal{F}}_1\left(m+p;p;\boldsymbol{\Omega}\left(\boldsymbol{I}_m+\boldsymbol{F}^{-1}\right)^{-1}\right)
\end{align}
from which we obtain upon using the transformation $\boldsymbol{V}=\left(\boldsymbol{I}_m+\boldsymbol{F}^{-1}\right)^{-1}$ with its differential form ${\rm d}\boldsymbol{F}=\det^{-2m}\left[\boldsymbol{I}_m-\boldsymbol{V}\right]{\rm d}\boldsymbol{V}$,
\begin{align}
h^{(0)}(\boldsymbol{V})=K_d(m,m,p)\text{etr}\left(-\boldsymbol{\Omega}\right){\det}^{p-m}[\boldsymbol{V}]{}_1\widetilde{\mathcal{F}}_1\left(m+p;p;\boldsymbol{\Omega}\boldsymbol{V}\right).
\end{align}
Now keeping in mind the relationship between the leading eigenvalue of $\boldsymbol{V}$, $v_{\max}$,  and the leading eigenvalue of $\boldsymbol{F}$, $\lambda_{\max}$, given by
\begin{align}
\label{vartrans}
    v_{\max}=\frac{\lambda_{\max}}{1+\lambda_{\max}},
\end{align}
for convenience, we may evaluate the c.d.f. of $v_{\max}$. To this end, following \cite{muirhead,mathaiB,koev,dumit}, we may write the c.d.f. of $v_{\max}$ as
\begin{align}
    F_{v_{\max}}(t)=\Pr\left\{v_{\max}\leq t\right\}=\Pr\left\{\boldsymbol{V}\prec t\boldsymbol{I}_m\right\}
\end{align}
where $\boldsymbol{V}\prec t\boldsymbol{I}_m$ implies the region of $t$ for which the matrix $t\boldsymbol{I}_m-\boldsymbol{V}$ is positive definite. As such, the desired c.d.f. can be expressed as 
\begin{align}
    F_{v_{\max}}(x)=K_d(m,m,p)\text{etr}\left(-\boldsymbol{\Omega}\right)\int_{\boldsymbol{0}}^{t\boldsymbol{I}_m}{\det}^{p-m}[\boldsymbol{V}]\;{}_1\widetilde{\mathcal{F}}_1\left(m+p;p;\boldsymbol{\Omega}\boldsymbol{V}\right) {\rm d}\boldsymbol{V}
\end{align}
which simplifies, upon introducing the variable transformation $\boldsymbol{V}=t\boldsymbol{Y}$ with ${\rm d}\boldsymbol{V}=t^{m^2}{\rm d}\boldsymbol{Y}$, giving
\begin{align}
\label{cdfv}
 F_{v_{\max}}(t)&=K_d(m,m,p)\text{etr}\left(-\boldsymbol{\Omega}\right)t^{mp}\int_{\boldsymbol{0}}^{\boldsymbol{I}_m}{\det}^{p-m}[\boldsymbol{Y}]\;{}_1\widetilde{\mathcal{F}}_1\left(m+p;p;t\boldsymbol{\Omega}\boldsymbol{Y}\right) {\rm d}\boldsymbol{Y}\\
 &=\text{etr}\left(-\boldsymbol{\Omega}\right)t^{mp}
{}_0\widetilde{\mathcal{F}}_0\left(t\boldsymbol{\Omega}\right)\nonumber
\end{align}
where the second equality is due to \cite[Eq. 6.1.21]{Mathai}. Finally, noting that ${}_0\widetilde{\mathcal{F}}_0\left(t\boldsymbol{\Omega}\right)=\text{etr}\left(t\boldsymbol{\Omega}\right)=e^{t\omega }$ and $\text{etr}\left(-\boldsymbol{\Omega}\right)=e^{-\omega}$, we make use of the the variable transformation with (\ref{vartrans}) to yield 
\begin{align}
    F^{(0)}_{\lambda_{\max}}(t;\omega)=e^{-\frac{\omega}{1+t}}\left(\frac{t}{1+t}\right)^{m(\beta+m)}
\end{align}
thereby concluding the proof.

\section{Proof of Proposition \ref{pmatint}}\label{appc}
Following a similar approach as before, the density of $\boldsymbol{V}=\left(\boldsymbol{I}_m+\boldsymbol{F}^{-1}\right)^{-1}$ can be written as
\begin{align}
    h^{(\alpha)}(\boldsymbol{V})=K_d(m,n,p)\text{etr}\left(-\boldsymbol{\Omega}\right){\det}^{p-m}[\boldsymbol{V}]
    {\det}^{n-m}[\boldsymbol{I}_m-\boldsymbol{V}]{}_1\widetilde{\mathcal{F}}_1\left(n+p;p;\boldsymbol{\Omega}\boldsymbol{V}\right)
\end{align}
from which, in view of the same reasoning which led to (\ref{cdfv}), we obtain
\begin{align}
 F_{v_{\max}}(z)=K_d(m,n,p)\text{etr}\left(-\boldsymbol{\Omega}\right)z^{mp}\int_{\boldsymbol{0}}^{\boldsymbol{I}_m}{\det}^{p-m}[\boldsymbol{Y}]&
 {\det}^{n-m}[\boldsymbol{I}_m-z\boldsymbol{Y}]\nonumber\\
 & \qquad \times{}_1\widetilde{\mathcal{F}}_1\left(n+p;p;z\boldsymbol{\Omega}\boldsymbol{Y}\right) {\rm d}\boldsymbol{Y}.
\end{align}
Now let us assume that $\boldsymbol{\Omega}$ is rank one with $\text{tr}(\boldsymbol{\Omega})=\omega a\geq 0$. This in turn gives
\begin{align}
    F_{v_{\max}}(z)=K_d(m,n,p)e^{-\omega a}z^{mp}\int_{\boldsymbol{0}}^{\boldsymbol{I}_m}{\det}^{p-m}[\boldsymbol{Y}]&
 {\det}^{n-m}[\boldsymbol{I}_m-z\boldsymbol{Y}]\nonumber\\
 & \qquad \times{}_1\widetilde{\mathcal{F}}_1\left(n+p;p;z\boldsymbol{\Omega}\boldsymbol{Y}\right) {\rm d}\boldsymbol{Y}
\end{align}
which can be further simplified, keeping in mind that the matrix $\boldsymbol{\Omega Y}$ is rank-one, with the help of (\ref{eqhypdegenclassic}) to yield 
\begin{align}
    F_{v_{\max}}(z)=K_d(m,n,p)e^{-\omega a}z^{mp}\int_{\boldsymbol{0}}^{\boldsymbol{I}_m}{\det}^{p-m}[\boldsymbol{Y}]&
 {\det}^{n-m}[\boldsymbol{I}_m-z\boldsymbol{Y}]\nonumber\\
 & \qquad \times{}_1F_1\left(n+p;p;z\text{tr}\left(\boldsymbol{\Omega}\boldsymbol{Y}\right)\right) {\rm d}\boldsymbol{Y}.
\end{align}
To facilitate further analysis, we may replace the confluent hypergeometric function with its equivalent infinite series expansion and change the summation and integration operators to obtain

    \begin{align}
    F_{v_{\max}}(z)=K_d(m,n,p)e^{-\omega a}z^{mp}\sum_{\nu=0}^\infty
    \frac{(n+p)_\nu z^\nu}{(p)_\nu \nu !}\int_{\boldsymbol{0}}^{\boldsymbol{I}_m}{\det}^{p-m}[\boldsymbol{Y}]&
 {\det}^{n-m}[\boldsymbol{I}_m-z\boldsymbol{Y}]\nonumber\\
 & \qquad \times\text{tr}^\nu\left(\boldsymbol{\Omega}\boldsymbol{Y}\right) {\rm d}\boldsymbol{Y}.
\end{align}
Since $\boldsymbol{\Omega}$ can be factorized as $\boldsymbol{\Omega}=\omega a \boldsymbol{uu}^H$ with $\vert \vert \boldsymbol{u}\vert \vert=1$, we get
\begin{align}
\label{matintinexp}
    F_{v_{\max}}(z)=K_d(m,n,p)e^{-\omega a}z^{mp}\sum_{\nu=0}^\infty
    \frac{(n+p)_\nu }{(p)_\nu \nu !}z^\nu \omega^\nu\int_{\boldsymbol{0}}^{\boldsymbol{I}_m}{\det}^{p-m}[\boldsymbol{Y}]&
 {\det}^{n-m}[\boldsymbol{I}_m-z\boldsymbol{Y}]\nonumber\\
 & \qquad \times\text{tr}^\nu\left(\boldsymbol{A}\boldsymbol{Y}\right) {\rm d}\boldsymbol{Y}
\end{align}
where $\boldsymbol{A}=a\boldsymbol{uu}^H$. 

Now let us focus on obtaining an alternative expression for the c.d.f. of $v_{\max}$, as a power series of $\omega$, with the help of Theorem \ref{thmain}. To this end, noting the relation $v_{\max}=\lambda_{\max}/(1+\lambda_{\max})$, we obtain, after some algebraic manipulation, from Theorem \ref{thmain}
\begin{align}
    F_{v_{\max}}(z)=\mathcal{K}(\alpha,\beta,m)
    e^{-\omega a}z^{m(\alpha+\beta+m)}
    \det\left[e^{\omega a z}\sum_{k=i-1}^{\alpha}\mu_i(k) (\omega a z)^k \;\;\;\;\; \Psi_{i,j}\left(\frac{z}{1-z}\right)\right]_{\substack{i=1,2,\ldots,\alpha+1\\
    j=2,3,\ldots,\alpha+1}}
\end{align}
where
\begin{align*}
 \mu_i(k)&= \frac{(\alpha+1-i)!(\alpha+\beta+2m+i-2)!}{(\alpha-k)!(k+1-i)!(k+\beta+2m+i-2)!}.
\end{align*}
Since only the first column of the above determinant depends on $\omega$ and the maximum power of $\omega$ therein is $\alpha$, we can conveniently write the $\nu$th term of the Taylor expansion of the determinant about the point $\omega=0$ as
\begin{align}
c_\nu=\det\left[(az)^\nu\sum_{\ell=0}^{\min(\alpha,\nu)} \frac{\nu!}{(\nu-\ell)!} \mu_i(\ell) \;\;\;\;\; \Psi_{i,j}\left(\frac{z}{1-z}\right)\right]_{\substack{i=1,2,\ldots,\alpha+1\\
    j=2,3,\ldots,\alpha+1}}.
\end{align}
Noting that fact that $\mu_i(\ell)$ is non-zero for $i\leq \ell+1$, we rearrange the factors in view of (\ref{eqpoch}) with some algebraic manipulation to arrive at
\begin{align}
c_\nu=(az)^\nu\det\left[\xi_i \;\;\;\;\; \Psi_{i,j}\left(\frac{z}{1-z}\right)\right]_{\substack{i=1,2,\ldots,\alpha+1\\
    j=2,3,\ldots,\alpha+1}}.
\end{align}
Finally, we equate the coefficients of $\omega^\nu$ in (\ref{matintinexp}) to $c_\nu$ with some algebraic manipulation to conclude the proof.

\section{The evaluation of $\mathcal{Q}_m(t,y)$} \label{appd}
For convenience, let us rewrite $\mathcal{Q}_m(t,y)$ as
\begin{align}
\label{Qappendix}
    \mathcal{Q}_m(t,y)=\frac{t^{\alpha m}}{2^{\varepsilon_m}}\mathcal{R}^{(\alpha,\beta)}_m\left(\frac{2}{t}-1,2y-1\right)
\end{align}
where
	\begin{align}
    \label{mehtaourint}
	\mathcal{R}^{(a,b)}_{m}(x,y) = \int_{-1}^1\int_{-1}^1\ldots\int_{-1}^1 \prod_{j=1}^{m}(1+z_{j})^{b}&(x-z_{j})^{a}\left(y-z_{j}\right)  \Delta_{m}^{2}(\mathbf{z})\; {\rm d}z_1 {\rm d}z_2\ldots{\rm d}z_m,
	\end{align}
    and $a,b$ are non-negative integers. The above multiple integral can easily be evaluated with the help of   Andr\'eief-Heine identity (i.e., the continuous form of Cauchy–Binet formula)\cite{ref:chiani,couillet} as determinant of a square matrix of size $m$. However, our goal is to obtain a determinant of a square matrix the size of which depends on $a$. To this end,
    our main strategy is to start with a related integral given in \cite[eqs. 22.4.2, 22.4.11]{mehta} as
    \begin{align} \label{eqmehta}
	\int_{-1}^1\int_{-1}^1\ldots\int_{-1}^1 \prod_{j=1}^{m}(1+z_{j})^{b} &\prod_{i=1}^{a+1}(s_{i}-z_{j}) \Delta_{m}^{2}(\mathbf{z}) {\rm d}z_1 {\rm d}z_2\ldots{\rm d}z_m                          		\nonumber\\&\qquad \qquad \qquad \qquad   =  \frac{K_{b,m}}{\Delta_{a+1}(\mathbf{r})}\det\left[C_{m+i-1}(s_{j})\right]_{i,j=1,2,...,a+1}
	\end{align}
    where
    \begin{align*}
    K_{b,m}=2^{m(b+m)}\prod_{j=0}^{m-1}\frac{j!(j+1)!(b+j)!}{(b+m+j)!}
    \end{align*}
    and $ C_{k}(x) $ are monic\footnote{A polynomial in which the coefficient of the highest order term is $1$.} polynomials orthogonal with respect to the weight $ (1+x)^{b} $, over $ -1\leq x \leq 1 $. Since Jacobi polynomials are orthogonal with respect to the preceding weight, we use $ C_{k}(x) = 2^{k}\frac{k!(k+b)!}{(2k+b)!}P_{k}^{(0,b)}(x)$ in (\ref{eqmehta}) with some algebraic manipulation to obtain
    \begin{align}\label{eqmehta1}
	\int_{-1}^1\int_{-1}^1\ldots\int_{-1}^1 \prod_{j=1}^{m}(1+z_{j})^{b} &\prod_{i=1}^{a+1}(s_{i}-z_{j}) \Delta_{m}^{2}(\mathbf{z}) {\rm d}z_1 {\rm d}z_2\ldots{\rm d}z_m 
	 \nonumber\\
  & \qquad \qquad \qquad = \frac{\tilde{K}_{b,m}}{\Delta_{a+1}(\textbf{r})}\det\left[P_{m+i-1}^{(0,b)}(r_{j})\right]_{i,j=1,2,...,a+1}
	\end{align}
	where
	\begin{align*}
	\tilde{K}_{b,m} &= K_{b,m}\prod_{j=1}^{a+1}\frac{2^{m+j-1}(m+j-1)!(m+b+j-1)!}{(2m+2j+b-2)!}.
	\end{align*}
  In the above, $s_i$s are generally distinct parameters. Nevertheless, if we choose $s_i$ such that
  \begin{align*}
  s_i=\left\{\begin{array}{ll}
  y & \text{if $i=1$}\\
  x & \text{if $i=2,3,\ldots,\alpha+1$},
  \end{array}\right.
  \end{align*}
    then the left side of (\ref{eqmehta1}) coincides with the multidimensional integral of our interest in (\ref{mehtaourint}). Under the above parameterization, however, the right side of (\ref{eqmehta1}) assumes the indeterminate form $0/0$. Therefore, to circumvent this technical difficulty, capitalizing on an approach given in \cite{khatri, couillet}, instead of direct substitution, we use the following limiting argument
    \begin{align}\label{khatrilimitq}
	&\mathcal{R}^{(a,b)}_{m}(x,y) = \tilde{K}_{b,m}\;\lim_{\substack{s_{1}\to y\\s_{2}, s_{3} ,...,s_{a+1}\to x}}\frac{\det\left[P_{m+i-1}^{(0,b)}(s_{j})\right]_{i,j=1,2,...,a+1}}{\Delta_{a+1}(\mathbf{r})}
	\end{align}
 to yield %\cite{Khatri}, we write
 \begin{align}\label{eqkhatrilimit}
		\mathcal{R}^{(a,b)}_{m}(x,y)= \tilde{K}_{b,m}
        \frac{\det\left[P_{m+i-1}^{(0,b)}\left(y\right)\hspace{6mm} \displaystyle\frac{{\rm d}^{j-2}}{{\rm d}x^{j-2}}P_{m+i-1}^{(0,b)}(x)\right]_{\substack{i=1,2,...,a+1\\j=2,3,...,a+1}}}
        {\det\left[y^{i-1}\hspace{6mm} \displaystyle \frac{{\rm d}^{j-2}}{{\rm d}x^{j-2}}x^{i-1}\right]_{\substack{i=1,2,...,a+1\\j=2,3,...,a+1}}}.
	\end{align}
    Now the determinant in the denominator of (\ref{eqkhatrilimit}) admits
    \begin{align*}
    \det\left[y^{i-1}\hspace{6mm} \frac{{\rm d}^{j-2}}{{\rm d}x^{j-2}}x^{i-1}\right]_{\substack{i=1,2,...,a+1\\j=2,3,...,a+1}}=\prod_{j=1}^{a-1}j!\left(x-y\right)^{a},
    \end{align*}
    whereas the numerator can be evaluated with the help of (\ref{jacobiDerivative}) to yield
    \begin{align*}
    &\det\left[P_{m+i-1}^{(0,b)}\left(y\right)\hspace{6mm} \frac{{\rm d}^{j-2}}{{\rm d}x^{j-2}}P_{m+i-1}^{(0,b)}(x)\right]_{\substack{i=1,2,...,a+1\\j=2,3,...,a+1}}\\
    &\qquad  =2^{-\frac{a}{2}(a-1)}\det\left[P_{m+i-1}^{(0,b)}\left(y\right) \qquad (m+b+i)_{j-2}P_{m+i-j+1}^{(j-2,b+j-2)}(\omega_{x})\right]_{\substack{i=1,2,...,a+1\\j=2,3,...,a+1}}.
    \end{align*}
  Finally, substituting the above two expression into (\ref{eqkhatrilimit}) and then the result into (\ref{khatrilimitq}) gives
  \begin{align*}
		\mathcal{R}_{m}^{(a,b)}(x,y)&= \tilde{K}_{b,m}x^{a}2^{-\frac{a}{2}(a+1)}\left(\prod_{j=1}^{a-1}j!\right)^{-1}(x-y)^{-a}\\
		&\quad \qquad \qquad \times \det\left[P_{m+i-1}^{(0,b)}\left(y\right)\hspace{6mm}(m+i+b)_{j-2} P_{m+i-j+1}^{(j-2,b+j-2)}(x)\right]_{\substack{i=1,2,...,a+1\\j=2,3,...,a+1}}
	\end{align*}	
	which upon substituting into (\ref{Qappendix}) followed by the parameter change $a\to \alpha, b\to \beta, x\to \frac{2}{t}-1$, and $y\to 2y-1$ with some algebraic manipulation yields (\ref{qm}) .

%%%%%%
%% To balance the columns at the last page of the paper use this
%% command:
%%
%\enlargethispage{-1.2cm} 
%%
%% If the balancing should occur in the middle of the references, use
%% the following trigger:
%%
%\IEEEtriggeratref{4}
%%
%% which triggers a \newpage (i.e., new column) just before the given
%% reference number. Note that you need to adapt this if you modify
%% the paper.  The "triggered" command can be changed if desired:
%%
%\IEEEtriggercmd{\enlargethispage{-20cm}}
%%
%%%%%%

%%%%%%
%% References:
%% We recommend the usage of BibTeX:
%%
%\cite{atapattu2017exact}
\bibliographystyle{IEEEtran}

\bibliography{reference}
%%
%% where we here have assumed the existence of the files
%% definitions.bib and bibliofile.bib.
%% BibTeX documentation can be obtained at:
%% http://www.ctan.org/tex-archive/biblio/bibtex/contrib/doc/
%%%%%%

%% Or you use manual references (pay attention to consistency and the
%% formatting style!):
%\begin{thebibliography}{9}

\end{document}